\documentclass[a4paper,11pt]{article}
\usepackage{jheppub} 
\usepackage{lineno,amsmath,bm}
\usepackage{mathtools}
\usepackage{upgreek}
\usepackage{booktabs,tikz-cd}
\usepackage{bbm}
\usepackage{enumitem}
\newtheorem{theorem}{Theorem}[section]
\newtheorem{conj}[theorem]{Conjecture}

 \arxivnumber{2609.19904} 

\title{Conformal blocks and braiding matrices for RCFTs I: Virasoro 
minimal models}

\author[a,b]{Suresh Govindarajan,}
\author[c]{Aditya Jain,}
\author[a,b]{Akhila Sadanandan}
\author[a]{ and Prasanta K. Tripathy}
\affiliation[a]{Deparment of Physics,\\
Indian Institute of Technology Madras, Chennai 600036 India}
\affiliation[b]{Centre for Operator Algebras, Geometry, Matter and Spacetime, \\Indian Institute of Technology Madras, Chennai 600036 India}
\affiliation[c]{Asia Pacific Center for Theoretical Physics (APCTP)\\
Pohang, Gyeongsangbuk 37673, Korea}

\emailAdd{suresh@physics.iitm.ac.in}
\emailAdd{adityaj2807@gmail.com}
\emailAdd{akhila@physics.iitm.ac.in}
\emailAdd{prasanta@iitm.ac.in}

\abstract{We develop a systematic method for computing the braiding 
matrices of four-point functions in RCFTs. This is implemented for the 
case of Virasoro minimal models. The first few terms of the conformal 
block series are directly computed through the Shapovalov form. The 
Fuchsian (BPZ) ODEs for order three and higher have accessory parameters 
that are fixed by this direct computation. We thus bypass the derivation 
from the null-state condition which gets tedious for higher-level null 
states.  The braiding F-matrices are connection matrices between the 
Frobenius solutions at two singular points of the ODE. The F-matrices 
are computed numerically at first, agreeing with the results of 
Dotsenko--Fateev, which uses the Coulomb-gas formalism. We find that 
after a change of basis, the F-matrix is rendered unitary, which 
determines (products of) the three-point structure constants. We 
conjecture that the squares of the entries of the unitary F-matrix lie 
in a cyclotomic extension of the rational numbers. This enables us to 
convert our numerical estimates for the F-matrix into exact ones. The 
formalism is illustrated through numerous examples in Virasoro minimal 
models. We also discuss how these methods can be extended to cases 
involving symmetries that extend the Virasoro symmetry as well as for 
tenable examples that arise from the holomorphic modular bootstrap 
program.
}

\begin{document}
\maketitle
\flushbottom

\clearpage
\section{Introduction}

The holomorphic modular bootstrap aims to classify rational conformal 
field theories (RCFTs) by studying admissible solutions of modular 
linear differential equations (MLDEs) 
\cite{Mathur:1988na,Mathur:1988gt,Mathur:1989pk}. Admissible characters 
are solutions of an MLDE with non-negative integral $q$-series with the 
leading coefficient for the identity operator being unity. Recently, 
this approach received a boost when the same MLDE was used to obtain the 
exact S-matrix~\cite{Govindarajan:2026frs}. This enabled one to further 
study whether a given admissible solution had well-defined fusion rules. 
Such characters were called \textit{tenable}. With this in hand, all 
tenable solutions of MLDEs with one accessory parameter and up to six 
characters and central charge $\leq 24$ have been 
classified~\cite{Govindarajan:2026cgv}.

The algebraic content of an RCFT is encoded in a modular tensor category 
(MTC), which is associated with a topological field theory in $2+1$ 
dimensions~\cite{Turaev:1994,Rowell:2009}. Besides the modular data (the 
S- and T-matrices), an MTC is equipped with additional data, the 
braiding matrices, which are constrained by the pentagon and hexagon 
identities~\cite{Moore:1988qv}. The modular data of unitary MTCs with up 
to twelve primaries has been classified~\cite{Ng:2023wsc}, but the 
classification does not determine the braiding matrices. Further, it is 
known through an exotic counterexample that MTCs are not fixed by the 
modular data alone~\cite{Mignard:2021}. Thus, it seems useful to 
associate MTCs with tenable solutions by computing their braiding 
matrices.

In this paper, we begin a program to eventually compute four-point 
functions and the associated braiding matrices for tenable solutions. By 
braiding matrices, we mean the \textbf{F-matrix} (fusion or 
associativity matrix), which describes a change in the fusion channel, 
and the \textbf{R-matrix} (braiding matrix), which exchanges two 
adjacent fields. We develop a general formalism for determining the 
F-matrices,\footnote{We shall see that the R-matrices are fixed by the 
conformal weights alone, and are thus computationally trivial. The 
F-matrices carry all the calculational complexity.} and apply it to the 
Virasoro minimal models. Along with a direct computation of the first 
few terms of the four-point functions, we use an ODE approach to compute 
the F-matrices numerically, similar to the one used to obtain the 
S-matrix from the MLDE in~\cite{Govindarajan:2026frs}. The results are 
compared with those obtained by Dotsenko and 
Fateev~\cite{Dotsenko:1984ad,Dotsenko:1984nm} as a proof of concept of our methods.\footnote{An implementation of the Dotsenko--Fateev formulae, in Mathematica, has been provided by~\cite{Esterlis:2016psv}.} Braiding matrices and structure constants for Virasoro minimal models have been long known in the literature through several algebraic approaches: from the Coulomb-gas 
construction~\cite{Felder:1989wv, Felder:1989wr, Felder:1989hq}, from boundary 
CFT~\cite{Runkel:1998he,Runkel:1999dz}, and from quantum-group 
methods~\cite{Gepner:1993eg,Gepner:1994bb,Fuchs:1993xu, Teschner:1995ga}, where they appear as $6j$-symbols. In our formalism, we obtain them as connection matrices associated with the ODE for the conformal blocks. This serves as a benchmark for eventually setting up the computation for tenable solutions.

A paper that has been influential in our understanding of the 
computation of braiding matrices is that of Cheng, Gannon and 
Lockhart~\cite{Cheng:2020srs} (see also \cite{Mahanta:2022fvl}). Several 
methods were proposed by these authors and, in particular, they propose 
a hybrid method that combined the MLDE approach with the theory of 
vector-valued modular forms (VVMFs) as developed by Bantay and 
Gannon~\cite{Bantay:2007zz,Gannon:2013jua}. The solutions to the MLDE 
are the \textit{core} conformal blocks associated with the four-point 
function, written in terms of the modular parameter $\tau$. The 
F-matrices in this approach appear as monodromy matrices of the MLDE. A 
recent computation of the monodromy matrix from the MLDE 
\cite{Govindarajan:2026frs} suggested that the hybrid method may not be 
necessary; the braiding matrices can be obtained within the MLDE 
approach.

While the MLDE approach is viable in principle, the order of the 
differential equation grows rapidly, and it becomes impractical beyond 
low rank. With this in mind, we set out to study the Fuchsian ODEs 
obtained using the decoupling of null states in correlation functions 
i.e. the so-called BPZ equations~\cite{Belavin:1984vu}. Going from the 
null-state to the BPZ ODE is a straightforward but tedious calculation. 
The solution to these ODEs are the conformal blocks. Christe and 
Ravanini used such ODEs to classify RCFTs~\cite{Christe:1988xy}; their 
equations are organized by the number of exchanged primaries in a given 
correlator, whereas the equations we consider are of order set by the 
null level, and are written for the \textit{core blocks}. It is these 
ODEs which map to MLDEs after a suitable change of coordinate.

We bypass the tedious computation of the BPZ ODE from the null-state 
equation by computing several terms in the conformal block as a 
power-series in the cross-ratio directly from the Shapovalov form\footnote{Virasoro conformal blocks were also obtained through the inverse Shapovalov form in \cite{Fortin:2025asz}.} and 
using it to determine the Fuchsian ODE (with three singularities) for 
the conformal blocks~\cite[see Lecture 37]{Hori:2008}. After the ODE is 
fixed, the F-matrices then follow as connection matrices between the 
Frobenius bases at two regular singular points. A by-product of the 
construction is that imposing unitarity on the braiding matrices 
determines (products of) the three-point structure constants. The 
conformal data associated with the Virasoro minimal models is known in 
closed form through the Coulomb-gas representation of Dotsenko and 
Fateev~\cite{Dotsenko:1984nm,Dotsenko:1984ad}, which enables many 
consistency checks. Note that the R-matrices are simply determined by 
the half-monodromy phases fixed by the conformal weights.

The braiding matrices are genuinely finer than the modular data. A 
Fuchsian equation is determined by its local exponents alone only when 
it is of second order with three singular points. In every other case 
accessory parameters remain, and these carry the same information as the 
global monodromy, that is, as the braiding matrices themselves. There is 
no route therefore from the modular data alone to the braiding matrices, 
and the knowledge of the (truncated) conformal block series is precisely 
what breaks the circularity.

The method proposed in this paper consists of the following steps:
\begin{enumerate}[itemsep=0em]

 \item Use the characters of a tenable solution to work out the 
 decomposition of the associated modules into irreducible Virasoro 
Verma modules.
\item Pick a four-point function and determine its core conformal blocks 
to some order in the cross-ratio $z$. Since the fusion rules are known 
we know the primaries in the various channels. This is explained in 
Section~\ref{sec:Background}.
\item Determine the ODE from the known terms in the core-blocks. Once 
the ODE is fixed, the core blocks may be computed to any desired order. 
This is explained in Section~\ref{sec:ODEs}.
\item Compute the F-matrices numerically as connection matrices between 
the solutions of the ODE at two of its singular points, say $z=0$ and 
$z=1$.
\item Convert the numerical F-matrix into an exact one. Using the cyclotomic
conjecture \ref{conjecture} on the entries of the orthogonal F-matrix, we convert numerical 
F-matrices into exact ones.
\end{enumerate}
We implement this for the case of Virasoro minimal models as a proof of 
concept. Our conjecture is also a an important contribution as it appears to work
well beyond just the Virasoro minimal models. We also discuss how to handle more general cases with extended 
symmetry and illustrate it with an example. Our future plan is to apply 
this method to more extended RCFTs, in particular those based on affine 
Kac-Moody algebras, and ultimately to tenable solutions with no known 
RCFT realization. Braiding matrices for RCFTs with extended symmetry have been discussed via a quantum group interpretation in~\cite{Moore:1989ni, Reshetikhin:1989qg, Alvarez-Gaume:1988izd, Alvarez-Gaume:1988bek, Gomez:1989sw, Alvarez-Gaume:1989blj}.

After the introductory section, Section~\ref{sec:Background} provides a 
background for the problem and discusses the direct computation of 
correlation functions. Section \ref{sec:crossing} discusses crossing 
symmetries and braiding matrices. In particular, we show how to 
construct unitary F-matrices and propose Conjecture \ref{conjecture} on 
their entries. In Section~\ref{sec:ODEs}, we discuss ODEs for conformal 
blocks arising from null states as well as from the direct computation. 
In Section~\ref{sec:examples}, we illustrate our method with numerous 
examples. In Section~\ref{sec:extended}, we discuss how the direct 
computation can also be applied to RCFTs with symmetries that extend the 
Virasoro algebra. We conclude in Section~\ref{sec:conclusion} with a few 
remarks. The appendices contain notation, explicit Shapovalov matrices 
to level 4 and formulae for BPZ ODEs for core blocks at levels 2, 3 and 
4.

\paragraph{Note added:} After the manuscript was completed, an article 
was submitted to arXiv~\cite{ivanova2026} that has some overlap with our 
work. The paper studies the higher-order differential equations 
satisfied by Dotsenko--Fateev integrals for four-point functions 
involving the degenerate fields $\phi_{n,1}$. Their equations should 
coincide with the BPZ-type Fuchsian ODEs that we obtain in 
Section~\ref{sec:ODEs}. Our ODEs are not always of hypergeometric type. 
Their analysis does not address anything regarding the braiding 
matrices, which remain the distinguishing content of the present work. 

After the first version of this paper appeared, Lhoste, Qiao and
Yamazaki~\cite{Lhoste:2026qce} determined the bulk OPE coefficients of
all E-series Virasoro minimal models, unitary and non-unitary. In their
approach the chiral F-matrices are taken as input, in the form of
normalized products of $U_q(\mathfrak{sl}_2)$ $6j$ symbols, and the reduced crossing equations are
then solved for the OPE coefficients. By contrast, we obtain the F-matrices independently, as connection matrices of the Fuchsian ODE,
without assuming any quantum-group structure. In all cases we have
checked, our unitary F-matrices coincide with these normalized $6j$
symbols up to sign conventions for the primary fields. 

\section{Background}
\label{sec:Background}

A standard reference for the material discussed in this section is the 
book by DiFrancesco et al.~\cite{DiFrancesco:1997nk}. A nice reference 
are the handwritten lecture notes on CFT by Hori~\cite{Hori:2008}. A 
more recent review is the one by Ribault~\cite{Ribault:2016sla}.

\subsection{Correlation  functions in RCFT}

Let $\phi_i(z,\bar{z})$ denote the set of primary fields in a 
two-dimensional RCFT and let $(h_i,\bar{h}_i)$ denote the conformal 
weights of the primary fields. Let the fusion rules be given by
\begin{equation}
\phi_i \times \phi_j = \sum_p {N_{ji}}^p\ \phi_p,
\end{equation}
where the fusion coefficient ${N_{ji}}^p$ (which are either $0$ or $1$ 
for Virasoro minimal models) can be written in terms of the modular 
S-matrix using the Verlinde formula.

Using the state-operator correspondence, we define the state
\[
|\phi_i\rangle := \lim_{z\rightarrow0} \phi_i(z,\bar{z}) |0\rangle.
\]
The dual state is defined as
\[
\langle \phi_i | = \lim_{z\rightarrow\infty}  z^{2h_i}{\bar{z}}^{2\bar{h}_i}\ \langle 0|\  \phi_i(z,\bar{z}).
\]
Conformal invariance fixes the forms of the correlation functions. The 
two-point correlation function is
\begin{equation}
\langle \phi_1(z_1,\bar{z}_1) \phi_2(z_2,\bar{z}_2)\rangle  =(z_{12})^{-2h_1} (\bar{z}_{12})^{-2\bar{h}_1}\delta_{2\bar{1}},
\end{equation}
where this fixes the normalization of the operators. The chiral 
three-point function
\begin{equation}
    \langle \phi_1(z_1) \phi_2(z_2) \phi_3(z_3)\rangle \propto 
    \frac{1}{z_{12}^{h_1+h_2-h_3} z_{23}^{h_2+h_3-h_1} 
    z_{13}^{h_3+h_1-h_2}},
\end{equation}
is fixed up to an overall constant: the OPE coefficient. The chiral 
three-point function after fixing the $SL(2,\mathbb{C})$ invariance by 
moving the operators at $(z_k,z_j,z_p)$ to $(0,1,\infty)$ is
\[
\langle \phi_{p}| \phi_j(1) |\phi_k\rangle  :={c_{jk}}^{p} = c_{jk\,\bar{p}}.
\]
The correlation function is non-vanishing when the corresponding fusion 
matrix, ${N_{jk}}^{p}$, is non-zero. Similarly, conformal invariance 
fixes the chiral four-point function to take the following form:
\begin{equation}
\langle \phi_1(z_1) \phi_2(z_2) \phi_3(z_3)\phi_4(z_4)\rangle = 
\prod_{1\leq i<j\leq 4}(z_{ij})^{\mu_{ij}} \Phi(z)
\end{equation}
where $z_{ij}=(z_i-z_j)$ and the cross-ratio $z = 
\frac{z_{12}z_{34}}{z_{13}z_{24}}$. We have the additional conditions on 
the $\mu_{ij}$
\begin{align*}
\sum_{j<i} \mu_{ji} +\sum_{j>i}\mu_{ij} &= -2h_i \ ,\quad \forall i.
\end{align*}
These are only four conditions, and there is some freedom -- different 
choices only lead to a modification of the function of the cross-ratio.  
A symmetric choice is to set $\mu_{ij}=\frac{h}{3}-h_i-h_j$, where 
$h=\sum_i h_i$.

Another choice naturally arises as follows. Consider the non-chiral 
four-point correlator evaluated as follows:
\begin{equation}
\langle \phi_i|\ \phi_j(1) \phi_k(z,\bar{z})\ |\phi_l\rangle :=
\sum_{p} ({c_{ji}}^p)^* \, {c_{kl}}^p \,|\langle ijkl\rangle_p(z)|^2 ,
\end{equation}
where we assume that $|z|<1$ and $\langle ijkl\rangle(z)$ is the chiral 
conformal block defined in~\eqref{Fblockdef}. The index $p$ is the set 
of primaries for which the fusion matrices ${N_{kl}}^p$ and ${N_{ji}}^p$ 
are non-zero. The number of primaries, $n$, that appear in the sum is 
thus given by
\begin{equation}
n:=\sum_p {N_{kl}}^p{N_{ji}}^p,
\end{equation}
where we sum over all the primaries of the theory. The details of this computation, as well as an explicit formula, is given below.

\subsection{Direct computation of  conformal blocks}
\label{sec:blockcompute}

The operator product expansion (OPE) contains much more information than 
just the three-point structure constant, also called the OPE 
coefficient. It has information on the secondaries under the primary. 
Every secondary is in one-to-one correspondence with a partition 
$\lambda$. Let $\lambda=( \ell_s\geq \ell_{s-1}\geq \cdots \geq 
\ell_1\geq1)$ with $|\lambda|=\sum_a\ell_a$. Then $|\lambda|$ gives the 
level of the secondary. To each partition 
$\lambda=(\ell_1,\ell_2,\cdots,\ell_s)$, associate the operator
\[
\mathcal{L}_{-\lambda} := L_{-\ell_1}L_{-\ell_2}\cdots 
L_{-\ell_s},\qquad 
\mathcal{L}_{\lambda}:=\mathcal{L}_{-\lambda}^\dagger.
\]
Then, the secondary over the primary labeled $p$ given by the partition 
$\lambda$ at level $|\lambda|$ is given by
\[
|\phi_p^\lambda\rangle :=\phi_p^\lambda(0)|0\rangle = 
|\mathcal{L}_{-\lambda}\ |\phi_p\rangle .
 \] 
The corresponding operator is written as $\phi_p^\lambda(0)$.

The operator product expansion can be written as follows:
\begin{equation*}
\phi_{k}(z) \phi_l(0) = \sum_p z^{h_p-h_k-h_l}\ {c_{kl}}^p 
\left(\sum_{\lambda} \beta_{kl}^{p,\lambda} z^{|\lambda|}\ 
\phi_p^\lambda(0)\right).
\end{equation*}
This implicitly defines $\beta_{kl}^{p,\lambda}$. The operator version 
of the above statement is the following.
\begin{equation} \label{betadef}
\phi_{k}(z) |\phi_l\rangle = \sum_{p} z^{h_p-h_k-h_l}\ {c_{kl}}^p 
\left(\sum_{\lambda} \beta_{kl}^{p,\lambda} z^{|\lambda|}\ 
|\phi_p^\lambda\rangle\right).
\end{equation}

For some $\lambda'$ such that $|\lambda'|=|\lambda|$, taking the inner 
product of the above equation with $\langle \phi_p^{\lambda'}|$, we get 
the following three-point function of one secondary and two primaries:
\begin{equation}\label{3pta}
 \langle \phi_{p}^{\lambda'}| \phi_k(z) |\phi_l\rangle = {c_{kl}}^p\ 
 z^{h_p+|\lambda'|-h_k-h_l} \sum_{\lambda} 
 \mathcal{M}_{\lambda\lambda'}^{(p)}\ \beta_{kl}^{p,\lambda} ,
 \end{equation}
where we define the Shapovalov form by
 \[
 {\mathcal{M}^{(p)}_{\lambda\lambda'}}=\mathcal{M}^{(p)}_{\lambda'\lambda}= \langle\phi_p^\lambda|\phi_p^{\lambda'}\rangle .
 \]
This is a symmetric Gram matrix of size equal to the number of 
partitions of $|\lambda|$.

Another computation of the three-point function computed in~\eqref{3pta} 
is as follows.
\begin{equation}\label{3ptb}
\langle \phi^\lambda_{p}| \phi_k(z) |\phi_l\rangle =
\langle \phi_{p}| \mathcal{L}_\lambda\phi_k(z) |\phi_l\rangle  
= {c_{kl}}^p\ z^{h_p+|\lambda|-h_k-h_l}\ \alpha_{kl}^{p,\lambda},
\end{equation}
where
\[
 \alpha_{kl}^{p,\lambda}  =(z)^{-h_p+h_k+h_l-|\lambda|} \hat{L}_{\ell_s}\cdots \hat{L}_{\ell_1}\cdot (z)^{h_p-h_k-h_l},
\]
and $\hat{L}_\ell =z^{\ell+1}\partial_z + h_k(\ell+1)z^\ell$. An 
explicit formula is then given by
\begin{equation}\label{alphaformula}
\alpha_{kl}^{p,(\ell_1,\ldots,\ell_s)}= \prod_{x=1}^s \Big[(\ell_x+1)h_k 
+(h_p-h_k-h_l)+N_x\Big],
\end{equation}
with  $N_x =\sum_{i=1}^{x-1}\ell_i$.

Comparing~\eqref{3pta} and \eqref{3ptb}, for any two partitions 
$\lambda$ and $\lambda'$ with $|\lambda|=|\lambda'|$. we obtain the 
relation
\begin{equation}\label{alphabeta}
\alpha_{kl}^{p,\lambda} = \sum_{\lambda'} 
\mathcal{M}_{\lambda\lambda'}^{(p)}\ \beta_{kl}^{p,\lambda'}.
\end{equation}
In the above formula, we have explicit formulae for the 
$\alpha_{kl}^{p,\lambda}$ and can compute the Shapovalov matrix up to 
level 10 on a computer. Explicit form of the Shapovalov matrix is given 
in Appendix \ref{sec:shapmatrix} up to level 4. It is useful to think of 
$\alpha$ and $\beta$ for fixed $(k,l,p)$ as a column vector of size 
$|\lambda|$. Symbolically, one writes
\[
\alpha = \mathcal{M}_{|\lambda|}\cdot \beta.
\]
It is worth mentioning that the Shapovalov matrix is non-invertible in 
the presence of null vectors at level $|\lambda|$. As a consequence, 
$\beta$ is not unique given $\alpha$. The ambiguity in $\beta$ is given 
by a shift $v\in \ker {\cal M}_{|\lambda|}$, i.e. by a null descendant. 
Since $\cal M$ is symmetric, and $\alpha={\cal M}\cdot\beta$, we have 
$v^T \alpha=v^T {\cal M}\cdot \beta=({\cal M}v)^T\beta=0$, and the 
conformal block coefficient $\beta^T \alpha$ (defined in~\eqref{feqn}) 
is unchanged. Physically, the ambiguity amounts to adding a null state 
to the OPE, which does not affect the correlation function.

Let us compute the four-point function
\[
\langle \phi_i| \phi_j(1) \phi_k(z) |\phi_l\rangle.
\]
Using~\eqref{betadef} and the following 
\[
\langle \phi_i| \phi_j(1)= \left(\sum_{p,\lambda} {c_{ji}}^p\ 
\beta_{ji}^{p,\lambda}\ |\phi_{p}^{\lambda}\rangle \right)^\dagger ,
\]
we can compute the non-chiral four-point function by adding the 
anti-holomorphic part. We obtain
\begin{align}
\langle \phi_i| \phi_j(1) \phi_k(z) |\phi_l\rangle &= \sum_p 
(c_{ji}{}^p)^* \,c_{kl}{}^p\, \left| z^{h_p-h_k-h_l}\sum_{\lambda} 
(\beta_{ji}^{p,\lambda})^* \cdot \alpha_{kl}^{p,\lambda}\ 
z^{|\lambda|}\right|^2 .
\end{align}
In the above formula, we have assumed that $|z|$ is small and the right 
hand side is a formula valid for small enough $z$. In all our examples, 
the OPE coefficients as well as the $\alpha$ and $\beta$ matrices are 
real. We will hereafter drop the complex conjugation in subsequent 
expressions.

The four-point function is thus
\begin{equation} \label{Fblockdef}
\langle \phi_i| \phi_j(1) \phi_k(z) |\phi_l\rangle = \sum_{p} 
{c_{ij}}^p\, c_{kl}{}^p \, \left|\langle ijkl\rangle_p(z)\right|^2 ,
\end{equation}
where $\langle ijkl\rangle_p(z)$ is the chiral conformal block 
associated with the primary labeled $p$. In the $s$-channel, 
$z\rightarrow0$, limit, one has
\begin{equation}\label{feqn}
\langle ijkl\rangle_p(z) = z^{h_p-h_k-h_l}\sum_{\lambda} 
\beta_{ji}^{p,\lambda} \cdot \alpha_{kl}^{p,\lambda}\ z^{|\lambda|} = 
z^{h_p-h_k-h_l}\sum_{\lambda} \alpha_{ji}^{p,\lambda} \cdot 
\beta_{kl}^{p,\lambda}\ z^{|\lambda|} .
\end{equation}
The \textbf{core conformal block} (or simply, core block) of 
Cheng--Gannon--Lockhart~\cite{Cheng:2020srs} is given by a symmetric 
redefinition of the conformal block:
\begin{align} \label{coreblockdef}
[ijkl]_p(z) &:= z^{-\frac{h}3+h_k+h_l} \,(1-z)^{-\frac{h}3+h_j+h_k}\,\langle ijkl\rangle _p(z)  \nonumber\\
&= z^{h_p-\frac{h}3}\,(1-z)^{-\frac{h}3+h_j+h_k}\, \sum_{\lambda} \alpha_{ji}^{p,\lambda} \cdot \beta_{kl}^{p,\lambda}\ z^{|\lambda|},
\end{align}
where $h:=(h_i+h_j+h_k+h_l)$. Thus, the key to computing a conformal 
block is to compute the function
\[
\sum_{n=0}^\infty z^n \ \sum_{\lambda\vdash n} \alpha_{ji}^{p,\lambda} 
\cdot \beta_{kl}^{p,\lambda} .
\]
We wrote a program to compute the first 10 terms in the above 
expansion.\footnote{The Shapovalov matrix at level $|\lambda|$ has size 
$p(|\lambda|)$, the number of partitions of $|\lambda|$. At level ten 
this is a $42\times42$ matrix, and the size grows rapidly beyond that. 
Ten terms suffice for all the cases we consider.} This enables us to 
compute the core blocks to that order.  A different approach to 
computing four-point functions has been pursued by 
Perlmutter~\cite{Perlmutter:2015iya}.

\paragraph{Notation:} For the correlator $\langle 
\phi_1(\infty)\phi_2(1)\phi_3(z)\phi_4(0)\rangle$, we represent the 
chiral conformal block by $\langle1234\rangle$. Then, $[1234]$ will 
represent the corresponding core block. Since the core block is a 
symmetric redefinition of the chiral block, it is easy to see that there 
is some redundancy in the notation as $[1234]=[3412]=[2143]=[4321]$. 
Further, $\langle ijkl\rangle$ will indicate the vector of chiral blocks 
$\langle ijkl\rangle_p$, and $[ijkl]$ represents the vector of core 
blocks.

\section{Crossing symmetry and braiding matrices}
\label{sec:crossing}

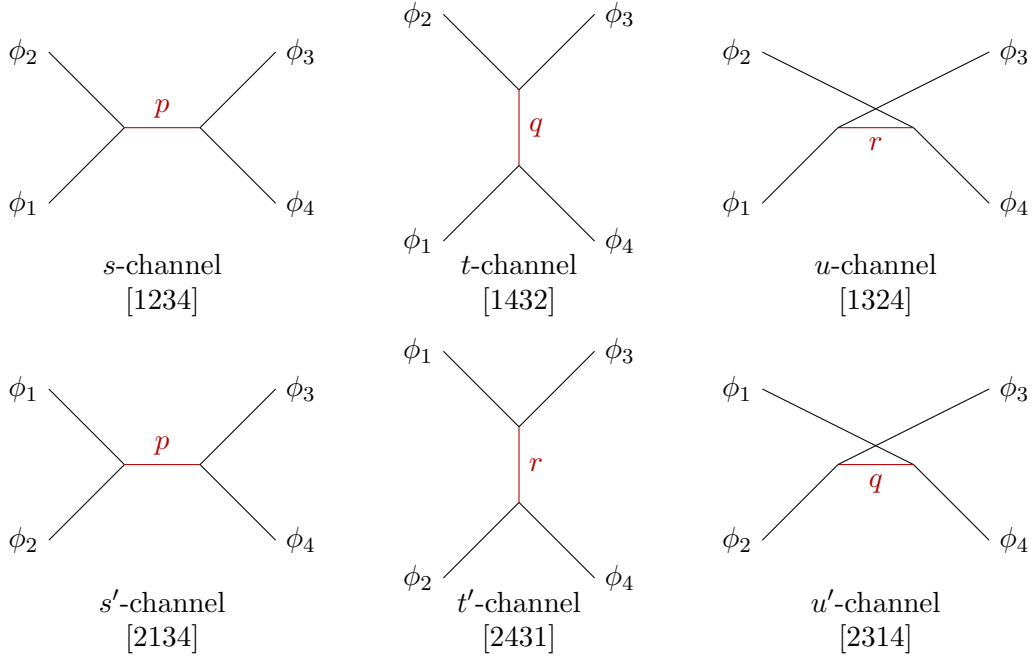
\begin{figure}[h!]
\centering
\begin{tikzpicture}[scale=1]
\draw (-1.5,1) -- (-0.5,0);
\draw (-1.5,-1) -- (-0.5,-0);
\draw[color=red!70!black] (-0.5,0) -- node[above] {$p$} (0.5,0);
\draw (0.5,0) -- (1.5,1);
\draw (0.5,-0) -- (1.5,-1);
\node[left] at (-1.5,1) {$\phi_2$};
\node[left] at (-1.5,-1) {$\phi_1$};
\node[right] at (1.5,1) {$\phi_3$};
\node[right] at (1.5,-1) {$\phi_4$};
\node at (0,-1.8) {$s$-channel};
\node at (0,-2.3) {$[1234]$};
\end{tikzpicture}
\qquad
\begin{tikzpicture}[scale=1]
\draw (1,-1.5) -- (0,-0.5);
\draw (-1,-1.5) -- (0,-0.5);
\draw[color=red!70!black] (0,-0.5) -- node[right] {$q$} (0,0.5);
\draw (0,0.5) -- (1,1.5);
\draw (0,0.5) -- (-1,1.5);
\node[right] at (1,-1.5) {$\phi_4$};
\node[left] at (-1,-1.5) {$\phi_1$};
\node[right] at (1,1.5) {$\phi_3$};
\node[left] at (-1,1.5) {$\phi_2$};
\node at (0,-1.8) {$t$-channel};
\node at (0,-2.3) {$[1432]$};
\end{tikzpicture}
\qquad
\begin{tikzpicture}[scale=1]
\draw (-1.5,1) -- (0.5,0);
\draw (-1.5,-1) -- (-0.5,-0);
\draw[color=red!70!black] (-0.5,0) -- node[below] {$r$} (0.5,0); 
\draw (-0.5,0) -- (1.5,1); 
\draw (0.5,0) -- (1.5,-1);
\node[left] at (-1.5,1) {$\phi_2$};
\node[left] at (-1.5,-1) {$\phi_1$};
\node[right] at (1.5,1) {$\phi_3$};
\node[right] at (1.5,-1) {$\phi_4$};
\node at (0,-1.8) {$u$-channel};
\node at (0,-2.3) {$[1324]$};
\end{tikzpicture}
\\
\begin{tikzpicture}[scale=1]
\draw (-1.5,1) -- (-0.5,0);
\draw (-1.5,-1) -- (-0.5,-0);
\draw[color=red!70!black] (-0.5,0) -- node[above] {$p$} (0.5,0);
\draw (0.5,0) -- (1.5,1);
\draw (0.5,-0) -- (1.5,-1);
\node[left] at (-1.5,1) {$\phi_1$};
\node[left] at (-1.5,-1) {$\phi_2$};
\node[right] at (1.5,1) {$\phi_3$};
\node[right] at (1.5,-1) {$\phi_4$};
\node at (0,-1.8) {$s'$-channel};
\node at (0,-2.3) {$[2134]$};
\end{tikzpicture}
\qquad
\begin{tikzpicture}[scale=1]
\draw (1,-1.5) -- (0,-0.5);
\draw (-1,-1.5) -- (0,-0.5);
\draw[color=red!70!black] (0,-0.5) -- node[right] {$r$} (0,0.5);
\draw (0,0.5) -- (1,1.5);
\draw (0,0.5) -- (-1,1.5);
\node[right] at (1,-1.5) {$\phi_4$};
\node[left] at (-1,-1.5) {$\phi_2$};
\node[right] at (1,1.5) {$\phi_3$};
\node[left] at (-1,1.5) {$\phi_1$};
\node at (0,-1.8) {$t'$-channel};
\node at (0,-2.3) {$[2431]$};
\end{tikzpicture}
\qquad
\begin{tikzpicture}[scale=1]
\draw (-1.5,1) -- (0.5,0);
\draw (-1.5,-1) -- (-0.5,-0);
\draw[color=red!70!black] (-0.5,0) -- node[below] {$q$} (0.5,0); 
\draw (-0.5,0) -- (1.5,1); 
\draw (0.5,0) -- (1.5,-1);
\node[left] at (-1.5,1) {$\phi_1$};
\node[left] at (-1.5,-1) {$\phi_2$};
\node[right] at (1.5,1) {$\phi_3$};
\node[right] at (1.5,-1) {$\phi_4$};
\node at (0,-1.8) {$u'$-channel};
\node at (0,-2.3) {$[2314]$};
\end{tikzpicture}
\caption{Various channels and the associated core blocks}\label{FigChannels}
\end{figure}

In a CFT, crossing symmetry tells us that the correlator of four 
distinct operators $\langle \phi_1\phi_2 \phi_3\phi_4 \rangle $ can be 
calculated in various ways. For instance, under the exchange 
$x_2\leftrightarrow x_4$, 
\begin{equation} 
\langle \phi_1 \phi_2 \phi_3 \phi_4 \rangle = \sum_{p} {c_{12}}^p 
\,c_{34}{}^p \, \langle 1234\rangle_p(z)\ \langle 1234\rangle_p(\bar{z}) =\sum_{q} 
c_{14}{}^q \,c_{23}{}^q \, \langle 1432\rangle_q(1-z)\ \langle 1432\rangle_q(1-\bar{z}) ,
\end{equation}
where $\langle 1234\rangle,\langle 1432\rangle$ represent the 
$s$-channel and $t$-channel conformal blocks, respectively. In terms of 
the core blocks, the prefactors cancel, and we get an identical 
equation:
\begin{equation} 
\sum_{p} {c_{12}}^p \,c_{34}{}^p \, [ 1234]_p(z)\ [ 1234]_p(\bar{z}) =\sum_{q} 
c_{14}{}^q \,c_{23}{}^q \, [1432]_q(1-z)\ [1432]_q(1-\bar{z}).
\end{equation}
The core blocks for the $s,t,u$-channels are thus $[1234],[1432], 
[1324]$, respectively. There are three other blocks leading to a total 
of six independent core blocks. We organize them as shown in the 
Figure~\ref{FigChannels}. The $s$- and $s'$-channels are related by an 
exchange of the first two operators, and the other primed channels 
follow simply.

\subsection{Braiding matrices from conformal blocks}

Crossing symmetry implies that the basis (of conformal blocks) for the 
$s$-channel and the $t$-channel must be related by a linear 
transformation. Similarly, the $s'$- and $t'$-channels as well as the 
$u$- and $u'$- channels are related by a linear transformation. Define 
the F-matrices that encode these linear transformations as follows:
\begin{gather}
\begin{gathered}\label{Fmatrices}
    [1432]_q(1-z) = \sum_p (F_{st})_q^p\ [1234]_p(z), \quad 
[2431]_r(1-z) =\sum_p (F_{s't'})_r^p\ [2134]_p(z),\\ 
[1324]_r(1-z) = \sum_q (F_{u'u})_r^q\ [2314]_q(z) .
\end{gathered}
\end{gather}
In each of the above relations, the two core blocks are obtained as 
solutions of a Fuchsian ODE (derived in Section~\ref{sec:ODEs}) at the 
regular singular points located at $z=0$ and $z=1$. Thus, the F-matrices 
are \textit{connection} matrices that connect the bases of solutions at 
the two singular points.

Similarly, define the R-matrices that encode the exchange of operators 
with the same intermediate primaries. For instance, the exchange of the 
first two operators, i.e. taking $[1234]$ to $[2134]$ does not change 
the $s$-channel OPE. In terms of the cross-ratio, this amounts to $ 
z\mapsto \frac{z}{z-1}, $ which fixes $z=0$, and swaps $z=1$ with 
$z=\infty$.\footnote{Equivalently, under the change of variable 
$z=\lambda(\tau)$ of Section~\ref{sec:MLDE} this is the modular 
transformation $\tau\to\tau+1$, under which 
$\lambda\to\lambda/(\lambda-1)$.} We therefore define
\begin{gather}
\begin{gathered}\label{Rmatrices}
[2134]_p(z) = \sum_{p'} (R_{ss'})_p^{p'}\, [1234]_{p'}\!\left(\tfrac{z}{z-1}\right),\qquad
[2314]_q(z) = \sum_{q'} (R_{tu'})_q^{q'}\, [1432]_{q'}\!\left(\tfrac{z}{z-1}\right),\\
[1324]_r(z) = \sum_{r'} (R_{t'u})_r^{r'}\, [2431]_{r'}\!\left(\tfrac{z}{z-1}\right).
\end{gathered}
\end{gather}
Because the intermediate primaries are unchanged ($p=p',q=q',r=r'$), 
each core block is mapped to itself up to a phase, which is determined 
by its leading behaviour in $z$, and the R-matrices are diagonal:
\begin{equation}
  (R_{ss'})_p = \varepsilon_p\, e^{i\pi (h_p-h/3)}, \quad
  (R_{tu'})_q = \varepsilon_q\, e^{-i\pi (h_q-h/3)}, \quad
  (R_{t'u})_r = \varepsilon_r\,e^{i\pi (h_r-h/3)} \ ,
\end{equation}
where $\varepsilon_p=\pm1$, $\varepsilon_q=\pm1$ and $\varepsilon_r=\pm1$ are signs to be fixed. We do not fix them here.
The inverse matrices simply give the inverse transformations. 

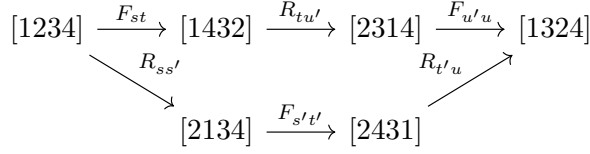
\begin{figure}[ht]
\centering
 \begin{tikzcd}
   { [1234]} \arrow[r,"F_{st}"]\arrow[rd,"R_{ss'}"] & {[1432]} \arrow[r,"R_{tu'}"] &  {[2314]}\arrow[r,"F_{u'u}"] & {[1324]} \\ 
& {[2134]} \arrow[r,"F_{s't'}"] & {[2431]} \arrow[ru,"R_{t'u}"] 
\end{tikzcd}
\caption{The hexagon identity}\label{hexagon}
\end{figure}

Combinations of the above six linear transformations enable us to relate 
the six-different channels. The hexagon identity then becomes a 
consistency condition which relates two distinct combinations of the six 
linear transformations. A pictorial proof is given in 
Figure~\ref{hexagon}. These are some of the braiding relations 
introduced by Moore and Seiberg \cite{Moore:1988qv, DeBoer:1990em}. The 
hexagon relation is satisfied by all the examples that we discuss in 
Section~\ref{sec:examples}.

Note that the determination of the F-matrices requires the knowledge of 
the differential equation satisfied by the conformal blocks, while the 
R-matrices are fixed almost entirely by the conformal weights, and are thus 
computationally trivial. Therefore, we shall not discuss R-matrices 
further, and F-matrices will be interchangeably referred to as braiding 
matrices.

\subsection{Permutation symmetry}
\label{sec:correlatorsymmetry}

So far, we have assumed that the operators in the four-point function 
are all distinct. However, if two or more of them are identical, then 
there is an additional (permutation) symmetry that reduces the number of 
independent core blocks, and some of the crossing channels 
($s,t,u,s',t',u'$) may get identified with each other. As a result, the 
number of unique braiding matrices decreases. For example, consider the 
case when two operators are identical, say $\phi_3=\phi_4$. Exchanging 
$3\leftrightarrow4$ must not change the correlator, and we obtain:
\begin{align*}
s:&\  (12)(34)\xrightarrow{\ 3\leftrightarrow4\ } (12)(43)=s',
\\
t:&\  (14)(32)\xrightarrow{\ 3\leftrightarrow4\ } (13)(42)=t',
\\
u:&\  (13)(24)\xrightarrow{\ 3\leftrightarrow4\ } (14)(23)=u'.
\end{align*}
Hence, $F_{st}=F_{s't'}$ describe the same braiding and along with 
$F_{u'u}$, we are left with \textit{two} independent F-matrices instead 
of three. There are only three independent core block vectors, viz. 
$[1233],[1323],[1332]$. Thus, the two F-matrices are
\begin{equation}
    [1332]_q(1-z) = \sum_p (F_{st})_q^p \,[1233]_p(z),\quad
    [1323]_r(1-z) =\sum_{r} (F_{uu})_r^{r} \,[1323](z).
\end{equation}
We present the number of independent core block vectors and F-matrices 
for all possible symmetries of the correlator:
\begin{center}
\begin{tabular}{c|ccccc}
type & $[1111]$ & $[1112]$ & $[1122]$ & $[1123]$ & $[1234]$\\\hline
\# core blocks & 1 & 1 & 3 & 3 & 6\\
\# F-matrices & 1 & 1 & 2 & 2 & 3
\end{tabular}
\end{center}
Thus, it suffices to study the following three types of correlators: all 
identical ([1111]), pairwise identical ([1122]), and all distinct 
([1234]).

\subsection{Orthogonal F-matrices and OPE coefficients}
\label{sec:unitaryFmatrix}

Consider the crossing symmetry equation
\begin{equation}\label{crossingcore}
\sum_p C_p\,[ijkl]_p(z)\ [ijkl]_p(\bar{z})\;=\; \sum_q 
C_q\,[ilkj]_q(1-z)\ [ilkj]_q(1-\bar{z}),
\end{equation}
where
\begin{equation}
C_p := {c_{ij}}^{p}\,{c_{kl}}^{p}, \qquad
C_q := {c_{il}}^{q}\,{c_{kj}}^{q},
\end{equation}
are products of the non-chiral three-point structure constants, or OPE coefficients. 
Similarly, define $C_r := {c_{ik}}^{r}\,{c_{jl}}^{r}$, for the 
$u$-channel. Substituting the definition~\eqref{Fmatrices} of the 
F-matrix into~\eqref{crossingcore}, crossing symmetries requires
\begin{equation}\label{crossingmatrix}
 F_{st}^{T} \cdot \mathrm{diag}(C_q)\cdot F_{st}= \mathrm{diag}(C_p) .
\end{equation}
Equivalently, defining
\begin{equation}\label{Fbardef}
\bar F_{st} := \mathcal{D}_p\, F_{st}\, \mathcal{D}_q^{-1}, \qquad
\mathcal{D}_p = \mathrm{diag}\big(\sqrt{C_p}\big), \quad
\mathcal{D}_q = \mathrm{diag}\big(\sqrt{C_q}\big),
\end{equation}
the relation~\eqref{crossingmatrix} reads
\begin{equation}
\bar F_{st}^{T}\cdot \bar F_{st} = \mathbbm{1},
\end{equation}
so that $\bar F_{st}$ is \emph{orthogonal}. Further, we see that the 
ratios of the elements of the F-matrix are related to the three-point 
structure constants:
\begin{align}
\frac{F_{pq}}{F_{qp}} = \sqrt\frac{C_p}{C_q}.
\end{align}
In the above formula, $p$ and $q$ are the indices of the F-matrix (the 
subscript \textit{st} has been removed for ease of presentation). These 
are $N(N+1)/2$ conditions among the $2N$ constants $C_p$ and $C_q$. 
Thus, they provide multiple consistency checks on the computation. For 
the other two F-matrices, one similarly writes
\begin{gather}
F_{s't'}^{T} \cdot \mathrm{diag}(C_r)\cdot F_{s't'}= \mathrm{diag}(C_p),\qquad
F_{u'u}^{T} \cdot \mathrm{diag}(C_r)\cdot F_{u'u} = \mathrm{diag}(C_q),
\end{gather}
and thus we obtain the orthogonal matrices
\begin{gather}
\bar F_{s't'} := \mathcal{D}_p\, F_{s't'}\, \mathcal{D}_r^{-1},\qquad
\bar F_{u'u} := \mathcal{D}_q\, F_{u'u}\, \mathcal{D}_r^{-1},
\end{gather}
where $\mathcal{D}_r = \mathrm{diag}\big(\sqrt{C_r}\big)$, of course. We emphasize that orthogonality is nothing other than crossing symmetry, 
rewritten. It is natural to compare the orthogonal matrices $\bar F$ 
with the F-matrices of an MTC, which are unitary for a unitary 
theory~\cite{Moore:1988qv,Rowell:2009}.

\subsection{A conjecture on the entries of the F-matrix}\label{sec:conj}

Unlike the S-matrix of an RCFT where there is a preferred basis, there 
is not one for the F-matrices. We fix this somewhat by demanding that 
the F-matrix be orthogonal. It is known that the entries of the normalized 
S-matrix i.e., $\bar{S}_{ij}=S_{ij}/S_{0j}$ are elements of the integer 
ring $\mathbb{Z}[\zeta(\bar{N})]$ of the cyclotomic extension of 
rationals, $\mathbb{Q}(\zeta_{\bar N})$, where ${\bar N}$ is the order 
of $T/T_{00}$~\cite{DeBoer:1990em,Ng:2007}. In particular, it implies 
that the quantum dimensions of operators, $d_i=S_{0i}/S_{00}$, lie in 
the integer ring $\mathbb{Z}[\zeta(\bar{N})]$.

Can we say something similar about the F-matrix? Consider the case of 
the unitary F-matrix for the Fibonacci MTC (that is realised as an RCFT 
by $G_{2,1}$) given by~\cite{Rowell:2009}
\[
\bar{F}=\begin{pmatrix}
\varphi^{-1} & \varphi^{-1/2}\\
 \varphi^{-1/2} &  -\varphi^{-1}
\end{pmatrix}, \quad \text{ with } \varphi=\frac{1+\sqrt5}{2}\in \mathbb{Z}[\zeta_5].
\]
However, $\sqrt{\varphi}=\bar{F}_{12}/\bar{F_{11}}$ does not belong to 
any cyclotomic extension of rationals but its square does.

Another example comes from the four-point correlator of four identical 
operators, $(1,3)$ in the Kac notation, for the minimal model 
$\mathcal{M}(k-1,k)$, with $k>5$. We can use the results of 
Dotsenko--Fateev~\cite[see Eq. (5.11)]{Dotsenko:1984ad} to compute the 
orthogonal F-matrix (in the basis ordered by the operators $(1,1)$, $(1,3)$ 
and $(1,5)$)
\[
\bar{F} = \small\begin{pmatrix}
\tfrac{1}{[3]} & \tfrac{1}{\sqrt{[3]}} & \tfrac{\sqrt{[5]}}{[3]}\\[3mm]
\tfrac{1}{\sqrt{[3]}} & \tfrac{[6]}{[3][4]} & -\tfrac{[2]\sqrt{[5]}}{[4]\sqrt{[3]}}\\[3mm]
\tfrac{\sqrt{[5]}}{[3]} & -\tfrac{[2]\sqrt{[5]}}{[4]\sqrt{[3]}} & \tfrac{[2]}{[3][4]}
\end{pmatrix}
\]
where $[n]=\sin(n\pi/ k)/\sin(\pi/ k)$ is a quantum integer. The 
square of all entries lie in $\mathbb{Q}(\zeta_{2k})$. In all the 
examples that we present in Section~\ref{sec:examples}, we observe that 
the squares of the entries of the orthogonal F-matrix lie in a cyclotomic 
extension of that rationals leading to the conjecture
\begin{conj}\label{conjecture}
The squares of the entries of the orthogonal F-matrix satisfy
\[
\left|\bar{F}_{ij}\right|^2 \in \mathbb{Q}(\zeta_{2\tilde{N}}),
\]
where $\tilde{N}$ is the least common multiple of the denominators of 
the exponents of the Fuchsian ODE at $z=0$ (see Section~\ref{sec:ODEs}).
\end{conj}

In the minimal model example considered above, the conjecture implies 
that the relevant field is $\mathbb{Q}(\zeta_{6k})$ and one sees that 
$\mathbb{Q}(\zeta_{2k})\subset \mathbb{Q}(\zeta_{6k})$. In all the 
examples that we studied in Section~\ref{sec:examples}, we find that the 
above conjecture holds. It helps us convert numerical results into exact 
ones.

In \cite{Kirillov:1990}, Kirillov and Reshetikhin (KR) computed the $6j$-symbols for
$U_q(sl(2))$. These are related to the F-matrices for the $su(2)$ WZW models at 
level $k$ with $q=\exp(2\pi i/(k+2))$ (see also~\cite{Alvarez-Gaume:1988izd}). Nawata et al. extended the KR result to multiplicity-free 6$j$-symbols of $U_q(sl(N))$~\cite{Nawata:2013ooa}.
The factorization of minimal model $\mathcal{M}(p,p')$ braid and fusion matrices into $U_q(sl(2))\otimes U_{q'}(sl(2))$ (with $q=e^{(2\pi i p/p')}$ and $q'=e^{(2\pi i p'/p)}$) 
data was shown by Alvarez-Gaum\'e et al.~\cite{Alvarez-Gaume:1988bek} and Felder et al.~\cite{Felder:1989wv}. These two results are compatible with our conjecture. We believe that it is valid for all RCFTs and not just Virasoro minimal models  or integrable representations of affine Kac-Moody algebras.

\textbf{Remark:} In examples where the fusion coefficients have multiplicities, one needs to find a suitable basis among the degenerate operators such that the conjecture holds. A weaker version of the conjecture is that $W_{pq}$ defined by
\[
W_{pq} := \sum_{\alpha\beta} C_{p\alpha,q\beta}\ C^{-1}_{q\beta,p\alpha} \in \mathbb{Q}(\zeta_M)\ ,
\]
for some $M$. Here $C_{p\alpha,q\beta}$ is the connection matrix where $p$ and $q$ are primary labels and $(\alpha,\beta)$ are the multiplicity labels. This quantity is invariant under changes of normalisations as well as basis changes within the multiplicity sectors. For affine Lie algebras, $M=2(k+g)$, where $k$ is the level and $g$ the dual Coxeter number.
There is a family of examples studied by Gu and Jockers where they determine the 6j symbols $U_q(sl(N))$ for a case with multiplicity 2~\cite{Gu:2014nva}. We have checked that their results are also compatible with the conjecture~\cite{affine}.

A related result due to Freedman and Wang who argue that the elements of 
the F-matrix belong to $\mathbb{Q}(\sqrt{d_1},\ldots,\sqrt{d_{n-1}})$ 
for a rank-$n$ MTC~\cite{Freedman:2006yr}. Our conjecture appears to be 
stronger than this.

\section{ODEs for conformal blocks}
\label{sec:ODEs}

In this section we will discuss how conformal blocks for four-point 
functions appear as solutions to ODE's in the cross-ratio. The original 
derivation makes use of the decoupling of null states which leads to the 
BPZ equation in the context of minimal models. As this method is tedious 
at higher orders, we use a combination of the first few terms in the 
four-point function and the exponents at various points to fix the ODE.

\subsection{ODEs from null states}

The presence of null states in the Verma module corresponding to a 
primary state $|\phi_1\rangle$ in Virasoro minimal models gives rise to 
the well known Belavin--Polyakov--Zamolodchikov (BPZ) equation. Let 
$|\chi^{(n)}_1\rangle = {\mathcal L}^{(n)}|\phi_1\rangle$ be a 
descendant at the $n^{\rm th}$-level of the primary $|\phi_1\rangle$, 
which is null. Then, the corresponding order-$n$ partial differential 
operator ${\mathcal L}^{(n)}(z)$ annihilates any correlator that 
contains the field $\phi_1(z)$, resulting in the respective BPZ 
equation. These BPZ equations become ordinary differential equations 
(ODEs) when expressed in terms of the core conformal blocks.

The ODEs at order-2 (and some specific examples at order-3) are well 
known in the literature~\cite{DiFrancesco:1997nk}. The general forms of 
these ODEs are outlined in Appendix~\ref{bpzsection}. In addition, we 
have derived the most general expression for the order-4 ODE from the 
null state condition at level four. As can be seen, the expressions are 
quite involved. In the following we consider the null state ODEs for 
several examples which we will study in later sections.

We will first consider the case where all the operators are identical. 
As an illustration, we take the minimal model $\mathcal{M}(6,5)$ (which 
describes the physics of the tetra-critical Ising model). The Verma 
module corresponding to the primary field $(1,2)$ (with weight $1/8$) 
has a null descendant at level two. Setting $h_1=h_2=h_3=h_4=1/8$ in 
\eqref{bpz2}, we find 
\begin{equation}
    \left[\dfrac{d^2}{dz^2} + \frac{2(2z-1)}{3z(z-1)} \dfrac{d}{dz} 
    - \frac{(z^2-z-1)}{12z^2(z-1)^2}\right]{\cal F}(z) = 0,
\end{equation}
where $\cal F$ represents the core block vector.

Alternatively, we can consider the primary field $(1,3)$, with conformal 
weight $2/3$, which has a null descendant at level three. We set 
$h_1=h_2=h_3=h_4=2/3$ in \eqref{bpz3} to obtain
\begin{align}\label{eq:bpz_order3_identical}
    \left[\dfrac{d^3}{dz^3} + \frac{2(2z-1)}{z(z-1)} \dfrac{d^2}{dz^2} 
    - \frac{2(2z^2-2z+29)}{27z^2(z-1)^2}\dfrac{d}{dz} + \frac{152(2z^3-3z^2-3z+2)}{729z^3(z-1)^3}\right]{\cal F}(z) = 0
\end{align}
Finally, before turning to pairwise identical operators, we take the 
primary field $(1,4)$ of weight $31/16$ in the minimal model 
${\mathcal{M}(12,11)}$. This admits a null descendant at level four. 
Setting $h_1=h_2=h_3=h_4=31/16$ in \eqref{bpz4}, we find the following 
order-4 ODE:
\begin{align}\label{eq:bpz_order4_identical}
    &\left[\dfrac{d^4}{dz^4} + \frac{4(2z-1)}{z(z-1)}\dfrac{d^3}{dz^3}
    -\frac{(119z^2-119z+407)}{24z^2(z-1)^2}\dfrac{d^2}{dz^2} \right.\cr
    & - \left.\frac{11(154z^3-231z^2+1101z-512)}{216z^3(z-1)^3}\dfrac{d}{dz}
    + \frac{155155(z^2-z+1)^2}{6912z^4(z-1)^4}\right] {\cal F}(z) = 0 
\end{align}

We now turn our attention to examples involving pairwise identical 
operators. First, consider pairs of operators with weights $1/8$ and 
$13/8$ in the tetra-critical Ising model. We obtain the following 
order-2 ODE:
\begin{equation}\label{eq:bpz_order2_pairwise}
    \left[\dfrac{d^2}{dz^2} + \frac{(7z-8)}{3z(z-1)} \dfrac{d}{dz}
    - \frac{(11z^2+7z-7)}{12z^2(z-1)^2} \right]{\cal F}(z) = 0.
\end{equation}

To conclude, we examine cases where all operators are distinct. First, 
consider the set of operators with weights $1/8$, $1/15$, $2/3$, and 
$1/40$ in the $\mathcal{M}(6,5)$ minimal model. The presence of the 
primary field with weight $1/8$ yields an order-2 ODE:
\begin{equation}
   \left[\dfrac{d^2}{dz^2} + \frac{(389z-187)}{180z(z-1)} \dfrac{d}{dz}
    - \frac{(29219z^2-34154z+8051)}{129600z^2(z-1)^2} \right]{\cal F}(z) = 0.
\end{equation}
As an aside, we note that this differential equation admits exact 
solutions that can be expressed in terms of elementary functions. 
Choosing the integration constants appropriately, we find their explicit 
forms to be:
\begin{align}
\begin{aligned}
& {\cal F}_1(z) = \frac{\big(1-\sqrt{z}\big)^{1/3} 
   +\big(1+\sqrt{z}\big)^{1/3}+3\sqrt{z}\left(\big(1-\sqrt{z}\big)^{1/3} 
   -\big(1+\sqrt{z}\big)^{1/3}\right)}{2z^{97/360}(1-z)^{41/180}},\\
   & {\cal F}_2(z) = \frac{3\left(\big(1-\sqrt{z}\big)^{1/3} 
   -\big(1+\sqrt{z}\big)^{1/3}+3\sqrt{z}\left(\big(1-\sqrt{z}\big)^{1/3} 
   +\big(1+\sqrt{z}\big)^{1/3}\right)\right)}{16z^{97/360}(1-z)^{41/180}}.
\end{aligned}
\end{align}

Our final example in this section is the core block involving distinct 
operators with weights $9/8, 11/5, 11/56,$ and $27/35$ in the 
$\mathcal{M}(10,7)$ minimal model. The primary field $(1,4)$ with weight 
$9/8$ possesses a null state at level four, yielding the following 
order-4 ODE:
\begin{align}
    &\left[{\phantom{+}}\dfrac{d^4}{dz^4} + \frac{(958z+121)}{210z(z-1)}\dfrac{d^3}{dz^3} 
    + \frac{(103636 z^2+1213964 z-1542467)}{117600 z^2 (z - 1)^2} \dfrac{d^2}{dz^2} \right.
  \cr &\left.  + \frac{ (180583192 z^3-676832388 z^2+1041292002 z -1979668403)}{148176000 z^3 (z-1)^3 } \dfrac{d}{dz} \right.
  \cr &  \left.-\frac{1}{
    497871360000 z^4(z-1)^4}\Big(634558562672 z^4-1883593099744 z^3 \right.
    \cr & \left.
    +1434178714728 z^2+1989032540744 z-2863475592889\Big)\right]{\cal F}(z) = 0.
\end{align}

Next, we understand how to determine the BPZ ODE without using the null 
state equation, and analyzing the parameters and behavior of the ODE at 
its singular points instead.

\subsection{ODEs from the direct computation}
\label{sec:direct}

A single four-point correlator may solve multiple BPZ ODEs corresponding 
to different nulls and we focus on the one with smallest order and call 
it $N$. An $N$-th order BPZ ODE is a Fuchsian ODE with regular singular 
points at $(0,1,\infty)$ and exponents $\rho^{\alpha}_i$ (with 
$\alpha\in(0,1,\infty)$, $i=1,\ldots,N$) and it satisfies the relation 
(see~\cite{Christe:1988xy} for a discussion):
\begin{equation}\label{eq:fuchs}
    \sum_{\alpha\in (0,1,\infty)} \sum_{i=1}^N \rho^{\alpha}_i=\frac{N(N-1)}{2},
\end{equation}
which is also called the \textit{Fuchs relation}.\footnote{Christe and 
Ravanini~\cite{Christe:1988xy} studied ODEs based on the conformal 
blocks ($\langle 1234\rangle(z)$ while our ODEs are those associated 
with the core blocks ($[1234]$). Thus, their exponents and subsequent 
analysis are different from ours.} Thus, the core block vector 
$[ijkl](z)$ solves the ODE:
\begin{align}
\left(\frac{d^{N}}{dz^{N}} + \sum_{a=1}^N \frac{p_a(z)}{z^a(1-z)^a}\, \frac{d^{N-a}}{dz^{N-a}}\right)[ijkl](z)=0\ ,
\end{align}
where $p_a(z)$ is a polynomial of degree $a$.

The order $N$ is determined by the null-state structure of the external 
operators, with $N\geq n$, where $n$ is the number of \textit{physical} 
intermediate primaries.\footnote{For Virasoro minimal models, the 
statement $N\geq n$ can be verified directly from the fusion rules 
\cite{Belavin:1984vu}.} We will focus, at first, on the case when the 
number of intermediate primaries is the same as the order. At each 
singular point $\alpha=\{0,1,\infty\}$, the indicial equation is a 
degree-$N$ polynomial in the Frobenius exponent giving rise to $N$ 
exponents $\rho_i^{(\alpha)}$:
\begin{align}
\rho_p^{(0)} = h_p - \frac{h}{3}\,,\quad
\rho_q^{(1)} = h_q - \frac{h}{3}\,,\quad
\rho_r^{(\infty)} = h_r - \frac{h}{3}\,,
\end{align}
where $p,q,r$ correspond to the intermediate primaries in the $s$-, 
$t$-, and $u$-channels, respectively (refer to Figure 
\ref{FigChannels}).

The number of parameters that appear in the ODE are $N(N+3)/2$. The 
local monodromy exponents uniquely fix the so-called \textit{rigid} 
parameters, and for an order-$N$ ODE,
\begin{align}
\text{number of rigid parameters}=(3N-1),
\end{align}
where we subtract the constraint coming from the Fuchs relation. Beyond 
these, the ODE has additional free parameters, called \textit{accessory} 
parameters. For order-$N$, one gets
\begin{align}
\text{number of accessory parameters}=\frac{(N-1)(N-2)}{2} .
\end{align}
These accessory parameters carry the same information as the 
\textit{global} monodromy, i.e. the braiding matrices. They are not 
fixed by the local exponents and require additional input. That input is 
the truncated core block series computed via the direct computation in 
Section \ref{sec:direct}:
\begin{align}
[ijkl]_p(z) = z^{h_p-h/3}\,(1-z)^{h_j+h_k-h/3}\,
  \sum_{m=0}^{L_{\max}} a_m^{(p)}\,z^m\,.
\end{align}

Since each physical block $[ijkl]_p$ satisfies the ODE by construction, 
substituting the series into the ODE ansatz and matching order by order 
in $z$ fixes all accessory parameters. Clearly, the number of series 
terms needed to fix the accessory parameters grows with $N$; in practice 
$L_{\max}\sim 10$ terms from the conformal block computation is 
sufficient for all cases studied here. With the ODE fully determined, 
its Frobenius recursion relation extends the power series of every 
solution to \emph{arbitrarily high order}, overcoming the practical 
limitation of the direct computation of the conformal block.

\subsection{Exceptional cases}
\label{sec:special}

For most of the four-point functions in Virasoro minimal models, we find 
$(N=n)$, all the exponents are non-integer-separated, and the Frobenius 
method produces $N$ independent power-series solutions (conformal 
blocks) at each singular point (OPE channel). Additionally, we encounter 
two exceptional situations, and in both cases the knowledge of the 
conformal block series provides the resolution.

\paragraph{Case 1: ODE order exceeds the number of physical blocks 
($N>n$).} The BPZ equation for a primary with null-state at level $N$ is 
an ODE of order $N$. It may happen that the number of exchanged 
primaries, i.e. physical conformal blocks is $n<N$, and the ODE has 
$N-n$ solutions that do not correspond to any conformal block. We call 
these `spurious solutions'. This case was also discussed 
in~\cite{Mukhi:2017ugw}, and also in the original modular bootstrap 
paper of Mathur--Mukhi--Sen~\cite{Mathur:1988na} in the context of 
MLDEs.

The Shapovalov algorithm gives us the $n$ physical conformal blocks 
explicitly. Substituting these into the order-$N$ ODE ansatz fixes all 
ODE parameters completely. Once the ODE is determined, its indicial 
equation at $z=0$ yields $N$ exponents: $n$ of them are the known 
physical exponents $\{h_i - h/3\}$, and the remaining $N-n$ spurious 
exponents are simply read off. It is then straightforward to compute the 
Frobenius solutions at all $N$ exponents by the ODE recursion.

The crucial point is that the F-matrix (connection matrix between the 
$z=0$ and $z=1$ bases) is \textit{upper triangular} in the basis where 
physical solutions are ordered before spurious ones. The spurious 
solutions do not mix with the physical intermediate channels: the 
$n\times n$ physical block of the connection matrix gives the braiding 
matrix directly, and the spurious entries are discarded.

One could alternatively employ the approach of Christe and 
Ravanini~\cite{Christe:1988xy}, and write an order-$n$ Fuchsian ODE with 
only the $n$ physical channels as solutions. However, such an ODE 
introduces additional apparent singularities beyond $\{0,1,\infty\}$, 
complicating the monodromy structure. We prefer to retain the 
three-point Fuchsian structure and work with the spurious solutions, 
since they anyway decouple from the physical braiding matrix.

\paragraph{Case 2: Exponents differing by a positive integer.} When two 
exponents at a given singular point differ by a positive integer, say 
$\rho_a^{(\alpha)}-\rho_b^{(\alpha)} =m\in\mathbb{Z}_{>0}$, the 
Frobenius solution encounters a degeneracy at order $z^m$, leaving the 
coefficient $a_m$ undetermined. In a generic Fuchsian ODE this 
`resonance' ambiguity can lead to a logarithmic solution. We do not 
expect that in the conformal blocks, since they are OPE expansions in 
integer powers of $z$.  The knowledge of the physical conformal block 
helps us resolves this discrepancy immediately. The core block for the 
intermediate primary $p$:
\begin{equation}
[ijkl]_p(z) = z^{\rho_p^{(0)}}
  \bigl(1 + a_1 z + \cdots + a_m z^m
  + \cdots\bigr),
\end{equation}
provides the coefficient $a_m$ at the discrepant order, making the 
system consistent, and the recursion continues normally to all higher 
orders.

\subsection{MLDEs for conformal blocks} 
\label{sec:MLDE}

The change of variable $z=\lambda(\tau)$ maps the conformal plane 
(sphere) to the upper half plane parametrized by $\tau$. The core blocks 
written in $\tau$ satisfy a modular differential equation, the MLDE. The 
order of the MLDE is equal to $x N$, where $N$ is the order of the ODE 
for a particular channel and $x=(1,3,6)$ depending on whether the 
operators are all identical, two are identical and all are distinct. 
Further, the core blocks for the various channels combine to form a 
vector-valued modular form (VVMF) of the full modular group. The 
F-matrices now arise as monodromies about a regular singular 
point~\cite{Cheng:2020srs,Mahanta:2022fvl}.

Let  $\mathbb{X}=\begin{pmatrix}
    [1234] & [2134] & [2314] & [1432] & [2431] & [1324]
\end{pmatrix}^T$. 
This is a VVMF of $PSL(2,\mathbb{Z})$. It is a $6N$ dimensional vector, 
as each core block has $N$ pieces. The modular S- and T- transformations 
of $X$ are given by the multiplier $\rho$ so that
\[
\rho(T) = \left(\begin{smallmatrix}
0 & R_{ss'} & 0 & 0 & 0 & 0 \\
R_{s's} & 0 & 0 & 0 & 0 & 0 \\
 0 &  0 & 0 &R_{u't} & 0 & 0 \\      
 0 &  0 &R_{tu'} & 0 & 0 & 0 \\
 0 & 0 & 0 & 0 & 0 & R_{t'u} \\
 0 & 0 & 0 & 0 &  R_{ut'} &0 
\end{smallmatrix}\right),
\quad\quad
\rho(S)= \left(\begin{smallmatrix}
0 & 0 & 0 & F_{st} & 0 & 0 \\
0 & 0 & 0 & 0 & F_{s't'} & 0 \\
 0 &  0 & 0 &0 & 0 & F_{u'u} \\      
 F_{ts} &  0 &0 & 0 & 0 & 0 \\
 0 & F_{t's'} & 0 & 0 & 0 & 0 \\
 0 & 0 & F_{uu'} & 0 &  0 &0 
\end{smallmatrix}\right).
\]
The entries are all $N\times N$ blocks. Note that $F_{ts}$ is the matrix 
inverse of $F_{st}$ and so on. Similarly, the R-matrices are such that 
$R_{s's}$ is the inverse of $R_{ss'}$ and so on.

The MLDE method of solving for the conformal blocks as well as the 
hybrid method combining the MLDE method with the theory of VVMFs was 
originally proposed by Cheng et al. in~\cite{Cheng:2020srs}. This works 
well when the rank of the modular form is $\leq 6$. It is impractical 
beyond that, and hence we did not pursue this. For instance, the rank 
becomes $24$ for distinct operators when $N=4$. Although impractical, it 
provides a global perspective on the problem of computing conformal 
blocks.

\section{Examples}
\label{sec:examples}

We now illustrate the formalism developed in the preceding sections 
through explicit examples in Virasoro minimal models ${\cal M}(p,q)= 
{\cal M} (q,p)$. For each example we carry out the full analysis: 
compute the conformal block series through the Shapovalov algorithm, 
read off the Fuchsian ODE parameters using the exponents (rigid) and the 
coefficients (accessory) of the blocks, determine the Frobenius 
solutions via ODE recursion, and extract the connection matrices. The 
resulting numerical F-matrices are then compared against the analytic 
expressions through the Dotsenko--Fateev formulae~\cite{Dotsenko:1984nm, 
Dotsenko:1984ad}. We then make a change of basis such that the 
F-matrices are unitary (orthogonal) and are able to give exact formulae 
using the conjecture on the entries, as described in 
Section~\ref{sec:unitaryFmatrix}. These unitary F-matrices are the ones 
usually given as a part of the MTC data.

The examples are organised by the symmetry of the operators in the 
correlator, which governs the number of independent F-matrices, as 
discussed in Section~\ref{sec:correlatorsymmetry}:
\begin{enumerate}[label=\arabic*.]
  \item \emph{All identical}: $\phi_1=\phi_2=\phi_3=\phi_4$
  \item \emph{Pairwise identical}: $\phi_1=\phi_2\neq\phi_3=\phi_4$
  \item \emph{All distinct}: $\phi_1 \neq\phi_2 \neq\phi_3\neq \phi_4$
\end{enumerate}
For each case, we present illustrative examples and numerically compute 
the braiding matrices. The examples marked with $\dagger$ belong to the 
exceptional cases studied in Section~\ref{sec:special}.

\subsection{All identical operators}
\label{sec:ex_identical}

For correlators involving four copies of the same operator, the $s$-, 
$t$-, $u$-channels are equivalent, thus carrying the same set of 
intermediate primaries $\{p\}$ and exponents. There is a single unique 
idempotent F-matrix: $F_{st}=F_{s't'}=F_{u'u}$.

\paragraph{Order-2:} The simplest example to consider is Ising CFT: 
minimal model ${\cal M}(4,3)$, and the correlator 
$\langle\sigma\sigma\sigma\sigma\rangle$. The primary operator $\sigma$ 
($h_\sigma=\frac{1}{16}$) contains a level-2 null, giving a second-order 
Fuchsian ODE. The fusion rule $\sigma\times \sigma=\mathbf{1} + 
\epsilon$, gives two conformal blocks, which are solutions of a 
second-order ODE. The Shapovalov algorithm produces the conformal 
blocks:
\begin{align}
\begin{aligned}
[{\sigma\sigma\sigma\sigma}]_{\mathbf{1}}(z)
&=z^{-\frac{1}{12}}\left(
1-
\tfrac{1}{24}z-
\tfrac{5}{1152}z^{2}+
\tfrac{161}{82944}z^{3}+
\tfrac{29305}{7962624}z^{4}+\cdots
\right),
\\
[{\sigma\sigma\sigma\sigma}]_{\epsilon}(z)
&=z^{\frac{5}{12}}\left(
1+
\tfrac{5}{24}z+
\tfrac{127}{1152}z^{2}+
\tfrac{6119}{82944}z^{3}+
\tfrac{438385}{7962624}z^{4}+\cdots
\right).
\end{aligned}
\end{align}
It is worth noting that the blocks above when written in the modular 
parameter $\tau$ through the Hauptmodul 
$\lambda(\tau)=\theta_{2}^{4}(\tau)/\theta_{3}^{4}(\tau)$ give the 
well-known characters of the $A_{1}$ Kac--Moody theory at level 1 
\cite{Cheng:2020srs}. The exponents at singularities $z=0,1,\infty$ are 
the same, and can be read off from the blocks: 
$\rho^{(0)}=\left\{-\frac{1}{12},\frac{5}{12}\right\}$. This completely 
determines the ODE that the blocks satisfy:
\begin{equation}
    \left[\dfrac{d^2}{dz^2} + \frac{(4z-2)}{3z(z-1)} \dfrac{d}{dz} 
    - \frac{-5(z^2-z+1)}{144z^2(z-1)^2}\right][{\sigma\sigma\sigma\sigma}](z) = 0.
\end{equation}
Employing the Frobenius method, one easily obtains the power series 
solutions for the above equation at $z=0$ and $z=1$. Comparing the 
solutions at $z=\frac12$, we obtain the connection matrix:
\begin{align}
F_{st}=\begin{pmatrix}
0.7071067804 & 0.3535533916 \\ 
1.414213562 & -0.7071067804
\end{pmatrix}.
\end{align}
This matches the analytic F-matrix a la Dotsenko--Fateev 
\cite{Dotsenko:1984nm} to arbitrary precision:
\begin{align}
F_{st}^{\rm DF}=\begin{pmatrix}
\frac{1}{\sqrt{2}} & \frac{\sqrt{2}\,\Gamma(\frac54)}{\Gamma(\frac14)}
\\
\frac{\Gamma(\frac14)}{2\sqrt{2}\,\Gamma(\frac54)} & -\frac{1}{\sqrt{2}}
\end{pmatrix}=
\begin{pmatrix}
\frac{1}{\sqrt{2}} & \frac{1}{2\sqrt{2}}
\\
\sqrt{2} & -\frac{1}{\sqrt{2}}
\end{pmatrix}.
\end{align}
We impose orthogonality on the product
\begin{align}
\bar F_{st}=\begin{pmatrix}
\sqrt{(c_{\sigma\sigma} {}^{\mathbf{1}} )^2} & 0
\\
0 & \sqrt{(c_{\sigma\sigma} {}^{\epsilon})^2}
\end{pmatrix}
\cdot
F_{st}
\cdot
\begin{pmatrix}
\sqrt{(c_{\sigma\sigma} {}^{\mathbf{1}} )^2} & 0
\\
0 & \sqrt{( c_{\sigma\sigma} {}^{\epsilon} )^2}
\end{pmatrix}^{-1}.
\end{align}
We obtain $c_{\sigma\sigma} {}^\epsilon=\frac12$, given that $c_{\sigma\sigma} {}^{\mathbf{1}}$ is normalized to 1. The orthogonal (unitary) F-matrix becomes
\[
\bar F_{st}={\frac{1}{\sqrt{2}}}
\begin{pmatrix}
1 & 1
\\
1 & -1
\end{pmatrix}.
\]
Observe that we are able to write an exact formula for the unitary 
F-matrix and have been able to determine the three-point structure 
constants as well. In more complicated cases, we will write out exact 
formulae for the unitary F-matrix. However, we will obtain numerical 
values for products of three-point structure constants. Obtaining 
individual structure constants requires working out several classes of 
four-point functions to have enough data for their estimation.

\paragraph{Order-2$^\dagger$:} We consider the Ising correlator 
$\langle\epsilon\epsilon\epsilon\epsilon \rangle$ since it displays one 
of the anomalous behaviors ($N>n$) described in 
Section~\ref{sec:special}. In particular, the primary operator 
$\epsilon$ ($h=\frac{1}{2}$) carries a null state at level-2, giving us 
a second-order ODE, while the fusion rule $\varepsilon\times\varepsilon= 
\mathbf{1}$ has a single intermediate and thus a single conformal block. 
This leads to only one of the solutions of the Fuchsian ODE being 
physical (corresponding to the $\mathbf{1}$ channel) and the other being 
spurious. We obtain
\begin{align}
[{\epsilon\epsilon\epsilon\epsilon}]_{\mathbf{1}}(z)=
z^{-\frac23}\left(
1-
\tfrac{1}{3}z-
\tfrac{8}{9}z^{2}+
\tfrac{49}{81}z^{3}+
\tfrac{125}{243}z^{4}+\cdots
\right).
\end{align}
When written using the Hauptmodul, this block produces the famous vacuum 
character of the $E_{8}$ theory at level 1 \cite{Cheng:2020srs}:
\begin{align}
[{\epsilon\epsilon\epsilon\epsilon}]_{\mathbf{1}}
\xrightarrow{z=\lambda(\tau)}
2^{-\frac43}q^{-\frac13}(1+248q+4124q^{2}+34752 q^{3}+213126q^{4}+\cdots).
\end{align}
We now employ the techniques discussed in Section~\ref{sec:special} to 
fix the ODE that the block satisfies, and obtain the second ``spurious" 
solution. The conformal block fixes the ODE as
\begin{equation}
    \left[\dfrac{d^2}{dz^2} + \frac{(4z-2)}{3z(z-1)} \dfrac{d}{dz} 
    - \frac{2(z^2-z+1)}{3z^2(z-1)^2}\right][{\epsilon\epsilon\epsilon\epsilon}](z) = 0,
\end{equation}
which gives us the exponents (identical for all the singularities) 
$\rho^{(\alpha)}=\{-\frac23,1\}$. One simply uses the Frobenius method 
to determine the power series solutions. Interestingly, the spurious 
second solution, when written in the Hauptmodul yields
\begin{align*}
q^{\frac12}\left(
1+\tfrac{228}{11}q+
\tfrac{34938}{187}q^{2}+
\tfrac{5163352}{4301}q^{3}+
\tfrac{764447229}{124729}q^{4}+\cdots
\right).
\end{align*}
This matches exactly with the second independent solution of the $E_{8}$ 
level-1 theory cited a few times in the literature \cite{Mathur:1988na, 
Gaberdiel:2007ve}. The rest of the computation is straightforward, and 
one gets the numerical F-matrix:
\begin{align}
F_{st}=\begin{pmatrix}
0.9999999716 & 7.478509175\times 10^{-8} \\ 
0.7605148849 & -0.9999999716
\end{pmatrix}\sim 
\begin{pmatrix}
1 & 0 \\
* & -1 
\end{pmatrix}\ .
\end{align}
The lower triangular nature of the matrix says that the solution of 
interest does not mix with the spurious solution.

\paragraph{Order-3$^\dagger$:} The example we consider resides in the 
minimal model ${\cal M}(6,5)$. We take the correlator of the (1,3) 
primary operator with $h=2/3$. The fusion rule $\phi_{\frac23}\times 
\phi_{\frac23}=\phi_{0}+\phi_{\frac23}+\phi_{3}$ gives us three 
conformal blocks, and the null occurs at level-3, making the BPZ 
equation third-order. The conformal blocks are
\begin{align}
\begin{aligned}
[\tfrac23,\tfrac23,\tfrac23,\tfrac23]_{0}(z)
&= z^{-\frac89}\left(
1
- \tfrac{4}{9}z
+ \tfrac{80}{81}z^{2}
+ \tfrac{1210}{2187}z^{3}
+ \tfrac{9617}{19683}z^{4}
+ \cdots
\right),
\\
[\tfrac23,\tfrac23,\tfrac23,\tfrac23]_{\tfrac23}(z)
&= z^{-\frac29}\left(
1
- \tfrac{1}{9}z
+ \tfrac{227}{324}z^{2}
+ \tfrac{1390}{2187}z^{3}
+ \tfrac{11951}{19683}z^{4}
+ \cdots
\right),
\\
[\tfrac23,\tfrac23,\tfrac23,\tfrac23]_{3}(z)
&= z^{\frac{19}{9}}\left(
1
+ \tfrac{19}{18}z
+ \tfrac{1121}{1053}z^{2}
+ \tfrac{30229}{28431}z^{3}
+ \tfrac{270875}{255879}z^{4}
+ \cdots
\right).
\end{aligned}
\label{eq:23block}
\end{align}
The exponents are $\rho^{(\alpha)}=\{-\frac89, -\frac29, 
\frac{19}{9}\}$. Note that the third-order Fuchsian ODE will have 
\textit{one} accessory parameter that the exponents cannot fix. The 
conformal block easily fixes it, and the ODE matches the BPZ equation 
in~\eqref{eq:bpz_order3_identical}.

It turns out that this example also belongs to the exceptional cases 
discussed in Section~\ref{sec:special}. The exponents $-\frac89 $ and 
$\frac{19}{9}$ differ by a positive integer: 
$\frac{19}{9}+\frac{8}{9}=3,$ and consequently, the coefficient $a_{3}$ 
for the power series solution at $\rho=-\frac{8}{9}$ cannot be 
determined through Frobenius recursion. This resonance discrepancy is 
resolved by the conformal blocks, since the first expression given 
in~\eqref{eq:23block} uniquely fixes the value of 
$a_{3}=\tfrac{1210}{2187}$. Following through, the numerical F-matrix is 
determined to be:
\begin{align}
F_{st}=\begin{pmatrix}
0.5000000015 & 0.6666666274 & 0.1604939435 \\ 
0.7499998296 & 2.830354\times 10^{-7} & -0.2407408613\\
1.557692060 & -2.076922631 & 0.4999997155
\end{pmatrix}.
\end{align}
The entries numerically match the analytic F-matrix:
\begin{align}
F_{st}^{\rm DF}&=
\begin{pmatrix}
\frac12	& \frac23 & \frac{13}{81} \\ 
\frac34 & 0 & -\frac{13}{54}\\
\frac{81}{52} & -\frac{27}{13} & \frac12
\end{pmatrix}.
\end{align}
Imposing unitarity, we obtain:
\[
c_{\frac23\frac23} {}^{\frac23}=\sqrt{\tfrac89},\quad
c_{\frac23\frac23} {}^{3}=\tfrac{26}{81},\qquad \bar{F}_{st}=\frac12\begin{pmatrix}
    1 & \sqrt2 & 1 \\
 \sqrt2 & 0 & -\sqrt2 \\
 1 & -\sqrt2 & 1
\end{pmatrix}.
\]

\paragraph{Order-4$^{\dagger}$:} We consider the correlator of the 
weight 31/16 primary in the minimal model 
$\mathcal{M}(12,11)$\footnote{This was the example studied by Cheng et 
al.~\cite{Cheng:2020srs}. Their computation needed additional inputs 
taken implicitly from Dotsenko-Fateev. We need no such inputs.}. The 
primary has a level-4 null state and fuses with itself to give four 
intermediate primaries with $h_p=\{0,\frac56,\frac72,8\}$. The core 
blocks are:
\begin{align}
\begin{aligned}
[\tfrac{31}{16},\tfrac{31}{16},\tfrac{31}{16},\tfrac{31}{16}]_{0}(z)
&= z^{-\frac{31}{12}}\left(
1
- \tfrac{31}{24}z
+ \tfrac{64945}{8064}z^{2}
- \tfrac{186589}{82944}z^{3}
+ \tfrac{40858465}{7962624}z^{4}
+ \cdots
\right),
\\
[\tfrac{31}{16},\tfrac{31}{16},\tfrac{31}{16},\tfrac{31}{16}]_{\tfrac{5}{6}}(z)
&= z^{-\frac{7}{4}}\left(
1
- \tfrac{7}{8}z
+ \tfrac{241045}{67456}z^{2}
+ \tfrac{233667}{539648}z^{3}
+ \tfrac{23720509995}{9515073536}z^{4}
+ \cdots
\right),
\\
[\tfrac{31}{16},\tfrac{31}{16},\tfrac{31}{16},\tfrac{31}{16}]_{\tfrac{7}{2}}(z)
&= z^{\frac{11}{12}}\left(
1
+ \tfrac{11}{24}z
+ \tfrac{29369}{40320}z^{2}
+ \tfrac{70397}{82944}z^{3}
+ \tfrac{42295661}{39813120}z^{4}
+ \cdots
\right),
\\
[\tfrac{31}{16},\tfrac{31}{16},\tfrac{31}{16},\tfrac{31}{16}]_{8}(z)
&= z^{\frac{65}{12}}\left(
1
+ \tfrac{65}{24}z
+ \tfrac{5767}{1152}z^{2}
+ \tfrac{648899}{82944}z^{3}
+ \tfrac{100783194515}{9069428736}z^{4}
+ \cdots
\right).
\end{aligned}
\end{align}
The exponents are $\rho^{\{\alpha\}}=\{-\frac{31}{12},-\frac74, 
\frac{11}{12},\frac{65}{12}\}$. There are \textit{three} accessory 
parameters in the order-4 ODE, which the blocks fix, and the Fuchsian 
ODE matches with~\eqref{eq:bpz_order4_identical}. The exponents turn out 
to have a resonance, i.e. differing by a positive integer 
($\frac{65}{12}-\frac{-31}{12}=8$). The conformal block for 
$\alpha=-\frac{31}{12}$ fixes the coefficient 
$a_8=\frac{47763761386595417}{5452690763022336}$. The numerical F-matrix 
is
\begin{align}
F_{st}=\small\begin{pmatrix}
0.2986824061 & 1.310428419 & 3.614370193 & 0.2415626115 \\
0.1862297970 & 0.5178583346 & -0.0003592140 & -0.1101089948 \\
0.09219160321 & 0.00002827040 & -0.7071536710 & 0.02726983541 \\
1.380732792 & -4.429991733 & 6.113682409 & -0.1093870697
\end{pmatrix},
\end{align}
and matches the analytic form. The unitary F-matrix gives us:
\[
\begin{aligned}
c_{\frac{31}{16}\frac{31}{16}}{}^{\frac56}&=2.65159815,\\
c_{\frac{31}{16}\frac{31}{16}}{}^{\frac72}&=6.26193341,\\
c_{\frac{31}{16}\frac{31}{16}}{}^{8}&=0.41814680,\\
\end{aligned}\quad
\bar F_{st}=\sqrt{\tfrac{2-\sqrt 3}{3}}\footnotesize
\begin{pmatrix}
1 & \sqrt{1+\sqrt3} & \sqrt{2+\sqrt3} & \sqrt{2+\sqrt3}\\
\sqrt{1+\sqrt3} & \sqrt3 & 0 & -\sqrt{2+2\sqrt3}\\
\sqrt{2+\sqrt3} & 0 & -\frac12({3+\sqrt3}) & \sqrt{\frac12(2+\sqrt3)}\\
\sqrt{2+\sqrt3} & -\sqrt{2+2\sqrt3} & \sqrt{\frac12(2+\sqrt3)} & \frac12 (1-\sqrt3)
\end{pmatrix}
\]
This is the F-matrix that was stated in~\cite{Cheng:2020srs}. We have 
also independently computed the same F-matrix using the MLDE method 
without using the hybrid method. The squares of the entries of $\bar{F}$ 
lie in $\mathbb{Q}(\zeta_{12})\subset \mathbb{Q}(\zeta_{24})$.

\subsection{Pairwise identical operators}
\label{sec:ex_pairwise}

For 4-point correlators containing two pairs of identical operators 
($h_{1}=h_{2}, h_{3}=h_{4}$), the $t$- and $u$-channels share the same 
intermediate primaries due to fusion of the same operators. There are 
two unique F-matrices, viz. $F_{st}$ ($=F_{s't'}$) and $F_{u'u}$.

\paragraph{Order-2:} We consider the $\mathcal{M}(6,5)$ minimal model 
and the correlator $\langle 
\phi_{\frac18}\phi_{\frac18}\phi_{\frac{13}8}\phi_{\frac{13}8} \rangle$. 
The fusion rules for the operators are
\begin{align}
\phi_{\frac18}\times \phi_{\frac18}=\phi_{0}+\phi_{\frac23}=\phi_{\frac{13}8}\times \phi_{\frac{13}8},\qquad \phi_{\frac18}\times \phi_{\frac{13}{8}}=\phi_{\frac23}+\phi_{3},
\end{align}
and the lowest null level is two, giving us an order-2 BPZ equation. The 
$s,t,u$-channel conformal blocks are, respectively,
\begin{align}
\begin{aligned}
[\tfrac{1}{8},\tfrac{1}{8},\tfrac{13}{8},\tfrac{13}{8}]_{0}(z)
&= z^{-\frac{7}{6}}\left(
1
- \tfrac{7}{12}z
+ \tfrac{445}{1152}z^{2}
+ \tfrac{6395}{41472}z^{3}
+ \tfrac{487091}{3981312}z^{4}
+ \cdots
\right),
\\
[\tfrac{1}{8},\tfrac{1}{8},\tfrac{13}{8},\tfrac{13}{8}]_{\tfrac{2}{3}}(z)
&= z^{-\frac{1}{2}}\left(
1
- \tfrac{1}{4}z
+ \tfrac{33}{256}z^{2}
+ \tfrac{115}{1024}z^{3}
+ \tfrac{23125}{229376}z^{4}
+ \cdots
\right),
\\
[\tfrac{1}{8},\tfrac{13}{8},\tfrac{13}{8},\tfrac{1}{8}]_{\tfrac{2}{3}}(z)
&= z^{-\frac{1}{2}}\left(
1
+ \tfrac{23}{16}z
+ \tfrac{273}{128}z^{2}
+ \tfrac{2275}{1024}z^{3}
+ \tfrac{37583}{16384}z^{4}
+ \cdots
\right),
\\
[\tfrac{1}{8},\tfrac{13}{8},\tfrac{13}{8},\tfrac{1}{8}]_{3}(z)
&= z^{\frac{11}{6}}\left(
1
+ \tfrac{31}{24}z
+ \tfrac{10997}{7488}z^{2}
+ \tfrac{6886663}{4313088}z^{3}
+ \tfrac{27032075}{15925248}z^{4}
+ \cdots
\right),
\\
[\tfrac{1}{8},\tfrac{13}{8},\tfrac{1}{8},\tfrac{13}{8}]_{\tfrac{2}{3}}(z)
&= z^{-\frac{1}{2}}\left(
1
- \tfrac{31}{16}z
+ \tfrac{165}{128}z^{2}
+ \tfrac{385}{1024}z^{3}
+ \tfrac{4103}{16384}z^{4}
+ \cdots
\right),
\\
[\tfrac{1}{8},\tfrac{13}{8},\tfrac{1}{8},\tfrac{13}{8}]_{3}(z)
&= z^{\frac{11}{6}}\left(
1
+ \tfrac{13}{24}z
+ \tfrac{3041}{7488}z^{2}
+ \tfrac{1454425}{4313088}z^{3}
+ \tfrac{60903095}{207028224}z^{4}
+ \cdots
\right).
\end{aligned}
\end{align}
Gathering the exponents $\rho^{(0)}=\{-\frac76,-\frac12\}$, 
$\rho^{(1)}=\rho^{(\infty)}=\{-\frac12, \frac{11}{6}\}$, it is easy to 
see that the Fuchsian ODE for the $s$-channel conformal blocks is given 
by~\eqref{eq:bpz_order2_pairwise}. The numerical F-matrices associated 
with the $s$-$t$ and $u'$-$u$ crossings are
\begin{align}
F_{st}=F_{s't'}={\small\begin{pmatrix}
0.4310611 & 0.23454854 \\
0.4060485 & -0.4418773
\end{pmatrix}},\quad
F_{u'u}=\small\begin{pmatrix}
0.5773502 & 0.6282944 \\
1.0610736 & -0.5773502
\end{pmatrix}.
\end{align}
Imposing unitarity, we find
\[
\sqrt{ c_{\frac{1}{8}\frac{1}{8}} {}^{\frac23}  \,c_{\frac{13}{8}\frac{13}{8}} {}^{\frac23} }=0.7506646607,\quad
c_{\frac{1}{8}\frac{13}{8}} {}^{\frac23}=0.5279399811,\quad
c_{\frac{1}{8}\frac{13}{8}} {}^{3}= 0.40625.
\]
And,
\[
\bar{F}_{st}=\bar{F}_{s't'} =\sqrt{\tfrac{1}{3}}{\small \begin{pmatrix}
     \sqrt{2} & 1 \\
 1 & -\sqrt{2}
\end{pmatrix}},\qquad
\bar{F}_{u'u}=\sqrt{\tfrac{1}{3}}{\small \begin{pmatrix}
     1 & \sqrt{2} \\
 \sqrt{2} & -1
\end{pmatrix}}
\]

\paragraph{Order-4:} Consider the pairwise identical correlator 
$\langle\phi_{\frac{1}{56}}\phi_{\frac{1}{56}}\phi_{\frac{5}{56}}\phi_{\frac{5}{56}}\rangle$ 
in the minimal model $\mathcal{M}(7,6)$. The relevant fusion rules are
\begin{align}
\begin{gathered}
    \phi_{\frac{1}{56}}\times \phi_{\frac{1}{56}}=\phi_{\frac{5}{56}}\times \phi_{\frac{5}{56}}=\phi_{0}+\phi_{\frac57}+\phi_{\frac43}+\phi_{\frac{1}{21}},\\
    \phi_{\frac{1}{56}}\times \phi_{\frac{5}{56}}=\phi_{\frac17}+\phi_{\frac{12}{7}}+\phi_{\frac{10}{21}}+\phi_{\frac{1}{21}}.
\end{gathered}
\end{align}
The conformal blocks are
\begin{align}
\begin{aligned}
[\tfrac{1}{56},\tfrac{1}{56},\tfrac{5}{56},\tfrac{5}{56}]_{0}(z)
&= z^{-\frac{1}{14}}\left(
1
- \tfrac{1}{28}z
- \tfrac{127}{9408}z^{2}
- \tfrac{675}{87808}z^{3}
+ \cdots
\right),
\\
[\tfrac{1}{56},\tfrac{1}{56},\tfrac{5}{56},\tfrac{5}{56}]_{\tfrac{5}{7}}(z)
&= z^{\frac{9}{14}}\left(
1
+ \tfrac{9}{28}z
+ \tfrac{43931}{238336}z^{2}
+ \tfrac{849335}{6673408}z^{3}
+ \cdots
\right),
\\
[\tfrac{1}{56},\tfrac{1}{56},\tfrac{5}{56},\tfrac{5}{56}]_{\tfrac{4}{3}}(z)
&= z^{\frac{53}{42}}\left(
1
+ \tfrac{53}{84}z
+ \tfrac{1488653}{3245760}z^{2}
+ \tfrac{294530683}{817931520}z^{3}
+ \cdots
\right),
\\
[\tfrac{1}{56},\tfrac{1}{56},\tfrac{5}{56},\tfrac{5}{56}]_{\tfrac{1}{21}}(z)
&= z^{-\frac{1}{42}}\left(
1
- \tfrac{1}{84}z
- \tfrac{395}{56448}z^{2}
- \tfrac{71131}{14224896}z^{3}
+ \cdots
\right),
\\
[\tfrac{1}{56},\tfrac{5}{56},\tfrac{5}{56},\tfrac{1}{56}]_{\tfrac{1}{7}}(z)
&= z^{\frac{1}{14}}\left(
1
+ \tfrac{3}{56}z
+ \tfrac{85}{3136}z^{2}
+ \tfrac{63769}{3512320}z^{3}
+ \cdots
\right),
\\
[\tfrac{1}{56},\tfrac{5}{56},\tfrac{5}{56},\tfrac{1}{56}]_{\tfrac{12}{7}}(z)
&= z^{\frac{23}{14}}\left(
1
+ \tfrac{79}{96}z
+ \tfrac{1213}{1792}z^{2}
+ \tfrac{24631}{43008}z^{3}
+ \cdots
\right),
\\
[\tfrac{1}{56},\tfrac{5}{56},\tfrac{5}{56},\tfrac{1}{56}]_{\tfrac{10}{21}}(z)
&= z^{\frac{17}{42}}\left(
1
+ \tfrac{349}{1680}z
+ \tfrac{542371}{4798080}z^{2}
+ \tfrac{37535795}{483646464}z^{3}
+ \cdots
\right),
\\
[\tfrac{1}{56},\tfrac{5}{56},\tfrac{5}{56},\tfrac{1}{56}]_{\tfrac{1}{21}}(z)
&= z^{-\frac{1}{42}}\left(
1
+ \tfrac{1}{24}z
+ \tfrac{1075}{44352}z^{2}
+ \tfrac{110525}{6386688}z^{3}
+ \cdots
\right),
\\
[\tfrac{1}{56},\tfrac{5}{56},\tfrac{1}{56},\tfrac{5}{56}]_{\tfrac{1}{7}}(z)
&= z^{\frac{1}{14}}\left(
1
+ \tfrac{1}{56}z
+ \tfrac{25}{3136}z^{2}
+ \tfrac{17431}{3512320}z^{3}
+ \cdots
\right),
\\
[\tfrac{1}{56},\tfrac{5}{56},\tfrac{1}{56},\tfrac{5}{56}]_{\tfrac{12}{7}}(z)
&= z^{\frac{23}{14}}\left(
1
+ \tfrac{551}{672}z
+ \tfrac{25325}{37632}z^{2}
+ \tfrac{1196813}{2107392}z^{3}
+ \cdots
\right),
\\
[\tfrac{1}{56},\tfrac{5}{56},\tfrac{1}{56},\tfrac{5}{56}]_{\tfrac{10}{21}}(z)
&= z^{\frac{17}{42}}\left(
1
+ \tfrac{331}{1680}z
+ \tfrac{506263}{4798080}z^{2}
+ \tfrac{172249069}{2418232320}z^{3}
+ \cdots
\right),
\\
[\tfrac{1}{56},\tfrac{5}{56},\tfrac{1}{56},\tfrac{5}{56}]_{\tfrac{1}{21}}(z)
&= z^{-\frac{1}{42}}\left(
1
- \tfrac{11}{168}z
- \tfrac{8711}{310464}z^{2}
- \tfrac{5399125}{312947712}z^{3}
+ \cdots
\right).
\end{aligned}
\end{align}
The numerical F-matrices are 
\begin{align}
\begin{aligned}
F_{st}=F_{s't'}&=\footnotesize\begin{pmatrix}
0.3391534737 & 0.003139151314 & 0.08972876349 & 0.5846828062 \\
-3.348220469 & 0.02485253406 & -0.8858281157 & 4.628910141 \\
-24.61286020 & -0.2278127704 & 3.255873348 & 21.21564024 \\
0.6327740172 & -0.004696834620 & -0.08370551171 & 0.4374045994
\end{pmatrix},\\
F_{u'u}&=\footnotesize\begin{pmatrix}
0.2569450422 & -0.004285445765 & -0.06797913839 & 0.7981859442 \\
-62.37684322 & -0.2569450422 & 16.50284442 & 47.85731343 \\
-1.942382804 & 0.03239593997 & -0.2569450422 & 3.016953819 \\
0.6698002653 & 0.002759066483 & 0.08860347048 & 0.2569450422
\end{pmatrix}.
\end{aligned}
\end{align}
The unitary F-matrices are
\begin{align*}
\bar F_{st}=\bar F_{s't'}=\tfrac{1}{\sqrt{3+3\gamma_{1}}}\footnotesize\begin{pmatrix}
    1 & -\sqrt{ \gamma_{1}} & -\sqrt{2} &  \sqrt{2\gamma_{1}} \\
-\sqrt{\gamma_{1}} & -1 & \sqrt{2\gamma_1}
   & \sqrt{2} \\
-\sqrt{2} &  \sqrt{2\gamma_{1}} & -1 & \sqrt{2 \gamma_{1}} \\
\sqrt{2\gamma_{1}} & \sqrt{2} & \sqrt{2 \gamma_{1}} & 1 
\end{pmatrix} ,\quad
\bar F_{u'u}=\tfrac{1}{\sqrt{3+3\gamma_{2}}}\footnotesize\begin{pmatrix}
    1 & \sqrt{ \gamma_{2}} & \sqrt{2} &  \sqrt{2\gamma_{2}} \\
-\sqrt{\gamma_{2}} & 1 & -\sqrt{2\gamma_2}
   & \sqrt{2} \\
-\sqrt{2} &  -\sqrt{2\gamma_{2}} & 1 & \sqrt{2 \gamma_{2}} \\
\sqrt{2\gamma_{2}} & -\sqrt{2} & -\sqrt{2 \gamma_{2}} & 1 
\end{pmatrix},
\end{align*}
where
\[
\gamma_1=2\cos\tfrac{2\pi}{7},\quad
\gamma_2=2+4\cos\tfrac{2\pi}{7}+2\cos\tfrac{4\pi}{7}.
\]
The cyclotomic numbers $\gamma_1$ and $\gamma_2$ are real elements of 
$\mathbb{Z}[\zeta_{7}]$ and the square of the entries of the 
$\bar{F}$-matrix lie in 
$\mathbb{Q}(\zeta_7)\subset\mathbb{Q}(\zeta_{42})$.

The (products of the) 3-point coefficients are numerically determined as
\begin{align*}
\begin{gathered}
\sqrt{c_{\frac{1}{56},\frac{1}{56}}^{\frac57}c_{\frac{5}{56},\frac{5}{56}}^{\frac57}}= 0.1131128,\ 
\sqrt{c_{\frac{1}{56},\frac{1}{56}}^{\frac43}c_{\frac{5}{56},\frac{5}{56}}^{\frac43}}= 0.0194871,\ 
\sqrt{c_{\frac{1}{56},\frac{1}{56}}^{\frac{1}{21}}c_{\frac{5}{56},\frac{5}{56}}^{\frac{1}{21}}}= 0.846432,\\
c_{\frac{1}{56}\frac{5}{56}}^{\frac17}=0.880555,\ 
c_{\frac{1}{56}\frac{5}{56}}^{\frac{12}{7}}=0.0072986,\ 
c_{\frac{1}{56}\frac{5}{56}}^{\frac{10}{21}}=0.0.164732,\ 
c_{\frac{1}{56}\frac{5}{56}}^{\frac{1}{21}}=0.961249.
\end{gathered}
\end{align*}

\subsection{All distinct operators}
\label{sec:ex_distinct}

The case of correlators containing distinct operators ($h_1\neq h_2\neq 
h_3\neq h_4$) is the richest one. The intermediate primaries propagating 
the $s,t,u$-channels respectively are all distinct, and there are three 
unique F-matrices.
    
\paragraph{Order-2:} We consider the correlator 
$\langle\phi_{\frac{1}{8}}\phi_{\frac23}\phi_{\frac{1}{40}}\phi_{\frac{1}{15}}\rangle$ 
in $\mathcal{M}(6,5)$. The fusion rules are
\begin{align*}
\begin{aligned}
\phi_{\frac{1}{8}}\times \phi_{\frac23}=\phi_{\frac{1}{40}}\times \phi_{\frac{1}{15}} = \phi_{\frac18}+\phi_{\frac{13}{8}},
\\
\phi_{\frac{1}{8}}\times \phi_{\frac{1}{15}}= \phi_{\frac23}\times \phi_{\frac{1}{40}}= \phi_{\frac{1}{40}}+\phi_{\frac{21}{40}},
\\
\phi_{\frac{1}{8}}\times \phi_{\frac{1}{40}}=\phi_{\frac23}\times \phi_{\frac{1}{15}}=\phi_{\frac{2}{5}}+\phi_{\frac{1}{15}}.
\end{aligned}
\end{align*}
There are 12 conformal blocks (2 for each of the $s,t,u,s',t',u'$-channels):
\begin{align*}
\begin{aligned}
[\tfrac{1}{8},\tfrac{2}{3},\tfrac{1}{40},\tfrac{1}{15}]_{\tfrac{1}{8}}(z)
&= z^{-\frac{61}{360}}\left(
1
- \tfrac{7}{40}z
- \tfrac{12311}{259200}z^{2}
- \tfrac{18492181}{839808000}z^{3}
+ \cdots
\right),
\\
[\tfrac{1}{8},\tfrac{2}{3},\tfrac{1}{40},\tfrac{1}{15}]_{\tfrac{13}{8}}(z)
&= z^{\frac{479}{360}}\left(
1
+ \tfrac{79}{120}z
+ \tfrac{128689}{259200}z^{2}
+ \tfrac{1012833557}{2519424000}z^{3}
+ \cdots
\right),
\end{aligned}
\end{align*}\vspace{-1em}
\begin{align*}
\begin{aligned}
[\tfrac{1}{8},\tfrac{1}{15},\tfrac{1}{40},\tfrac{2}{3}]_{\tfrac{1}{40}}(z)
&= z^{-\frac{97}{360}}\left(
1
+ \tfrac{221}{360}z
+ \tfrac{161203}{777600}z^{2}
+ \tfrac{110028023}{839808000}z^{3}
+ \cdots
\right),
\\
[\tfrac{1}{8},\tfrac{1}{15},\tfrac{1}{40},\tfrac{2}{3}]_{\tfrac{21}{40}}(z)
&= z^{\frac{83}{360}}\left(
1
+ \tfrac{163}{1080}z
+ \tfrac{71803}{777600}z^{2}
+ \tfrac{56973323}{839808000}z^{3}
+ \cdots
\right),
\end{aligned}
\end{align*}\vspace{-1em}
\begin{align*}
\begin{aligned}
[\tfrac{1}{15},\tfrac{2}{3},\tfrac{1}{40},\tfrac{1}{8}]_{\tfrac{2}{5}}(z)
&= z^{\frac{19}{180}}\left(
1
- \tfrac{1}{45}z
+ \tfrac{559}{259200}z^{2}
+ \tfrac{1052879}{139968000}z^{3}
+ \cdots
\right),
\\
[\tfrac{1}{15},\tfrac{2}{3},\tfrac{1}{40},\tfrac{1}{8}]_{\tfrac{1}{15}}(z)
&= z^{-\frac{41}{180}}\left(
1
- \tfrac{203}{360}z
- \tfrac{53471}{259200}z^{2}
- \tfrac{35398907}{279936000}z^{3}
+ \cdots
\right),
\end{aligned}
\end{align*}\vspace{-1em}
\begin{align*}
\begin{aligned}
[\tfrac{2}{3},\tfrac{1}{8},\tfrac{1}{40},\tfrac{1}{15}]_{\tfrac{1}{8}}(z)
&= z^{-\frac{61}{360}}\left(
1
+ \tfrac{1}{180}z
+ \tfrac{1781}{64800}z^{2}
+ \tfrac{2642423}{104976000}z^{3}
+ \cdots
\right),
\\
[\tfrac{2}{3},\tfrac{1}{8},\tfrac{1}{40},\tfrac{1}{15}]_{\tfrac{13}{8}}(z)
&= z^{\frac{479}{360}}\left(
1
+ \tfrac{121}{180}z
+ \tfrac{33221}{64800}z^{2}
+ \tfrac{131596109}{314928000}z^{3}
+ \cdots
\right),
\\
[\tfrac{2}{3},\tfrac{1}{15},\tfrac{1}{40},\tfrac{1}{8}]_{\tfrac{2}{5}}(z)
&= z^{\frac{19}{180}}\left(
1
+ \tfrac{23}{180}z
+ \tfrac{22051}{259200}z^{2}
+ \tfrac{2233727}{34992000}z^{3}
+ \cdots
\right),
\end{aligned}
\end{align*}\vspace{-1em}
\begin{align*}
\begin{aligned}
[\tfrac{2}{3},\tfrac{1}{15},\tfrac{1}{40},\tfrac{1}{8}]_{\tfrac{1}{15}}(z)
&= z^{-\frac{41}{180}}\left(
1
+ \tfrac{121}{360}z
+ \tfrac{36601}{259200}z^{2}
+ \tfrac{26526241}{279936000}z^{3}
+ \cdots
\right),
\\
[\tfrac{1}{15},\tfrac{1}{8},\tfrac{1}{40},\tfrac{2}{3}]_{\tfrac{1}{40}}(z)
&= z^{-\frac{97}{360}}\left(
1
- \tfrac{53}{60}z
- \tfrac{66017}{194400}z^{2}
- \tfrac{22789417}{104976000}z^{3}
+ \cdots
\right),
\\
[\tfrac{1}{15},\tfrac{1}{8},\tfrac{1}{40},\tfrac{2}{3}]_{\tfrac{21}{40}}(z)
&= z^{\frac{83}{360}}\left(
1
+ \tfrac{43}{540}z
+ \tfrac{349}{7200}z^{2}
+ \tfrac{3828143}{104976000}z^{3}
+ \cdots
\right).
\end{aligned}
\end{align*}
The numerically computed F-matrices are
\begin{align*}
F_{st}=\begin{pmatrix}
0.166667 & 0.7698 \\
2.92284 & -4.5
\end{pmatrix},
\ 
F_{s't'}=\begin{pmatrix}
0.529134 & 0.419974 \\
-5.35748 & 4.25223
\end{pmatrix},
\ 
F_{uu'}=\begin{pmatrix}
0.31498 & 0.839947 \\
-0.39685 & 1.05827
\end{pmatrix}.
\end{align*}
And their unitary versions turn out to be:
\begin{align*}
\bar F_{st}=\frac12\begin{pmatrix}
1 &  \sqrt{3}\\
\sqrt{3} & -1
\end{pmatrix},
\ 
\bar F_{s't'}=\frac{1}{\sqrt{2}}\begin{pmatrix}
1 & -1 \\
-1 & -1
\end{pmatrix},
\ 
\bar F_{u'u}=\frac{1}{\sqrt{2}}\begin{pmatrix}
1 & 1 \\
1 & -1
\end{pmatrix},
\end{align*}
The squares of the entries of the unitary F-matrices are elements of 
$\mathbb{Q}$. The values of (products of) the OPE coefficients obtained 
match the values given by Dotsenko--Fateev formalism.

\paragraph{Order-3:} We take the example of $\mathcal{M}(11,10)$ and the 
correlator 
$\langle\phi_{\frac{9}{11}}\phi_{\frac{21}{11}}\phi_{\frac{38}{11}}\phi_{\frac{60}{11}}\rangle$. 
There are 18 unique conformal blocks associated with the correlator:

\begin{align*}
\begin{aligned}
[\tfrac{9}{11},\tfrac{21}{11},\tfrac{38}{11},\tfrac{60}{11}]_{\tfrac{2}{11}}(z)
&= z^{-\frac{122}{33}}\left(
1
- \tfrac{259}{33}z
+ \tfrac{15407}{1089}z^{2}
- \tfrac{16884557}{754677}z^{3}
+ \cdots
\right),
\\
[\tfrac{9}{11},\tfrac{21}{11},\tfrac{38}{11},\tfrac{60}{11}]_{\tfrac{21}{11}}(z)
&= z^{-\frac{65}{33}}\left(
1
- \tfrac{719}{462}z
+ \tfrac{13967125}{10626462}z^{2}
- \tfrac{1190994872}{526009869}z^{3}
+ \cdots
\right),
\\
[\tfrac{9}{11},\tfrac{21}{11},\tfrac{38}{11},\tfrac{60}{11}]_{\tfrac{60}{11}}(z)
&= z^{\frac{52}{33}}\left(
1
+ \tfrac{163}{165}z
+ \tfrac{885881}{664290}z^{2}
+ \tfrac{110917549}{65764710}z^{3}
+ \cdots
\right),
\end{aligned}
\end{align*}\vspace{-1em}
\begin{align*}
\begin{aligned}
[\tfrac{9}{11},\tfrac{60}{11},\tfrac{38}{11},\tfrac{21}{11}]_{\tfrac{21}{11}}(z)
&= z^{-\frac{65}{33}}\left(
1
+ \tfrac{206}{231}z
- \tfrac{73979}{129591}z^{2}
- \tfrac{58579876}{12829509}z^{3}
+ \cdots
\right),
\\
[\tfrac{9}{11},\tfrac{60}{11},\tfrac{38}{11},\tfrac{21}{11}]_{\tfrac{60}{11}}(z)
&= z^{\frac{52}{33}}\left(
1
+ \tfrac{1907}{1320}z
+ \tfrac{96224071}{49157460}z^{2}
+ \tfrac{24694825423}{9733177080}z^{3}
+ \cdots
\right),
\\
[\tfrac{9}{11},\tfrac{60}{11},\tfrac{38}{11},\tfrac{21}{11}]_{\tfrac{119}{11}}(z)
&= z^{\frac{229}{33}}\left(
1
+ \tfrac{878}{231}z
+ \tfrac{208289}{22869}z^{2}
+ \cdots
\right),
\end{aligned}
\end{align*}\vspace{-1em}
\begin{align*}
\begin{aligned}
[\tfrac{60}{11},\tfrac{21}{11},\tfrac{38}{11},\tfrac{9}{11}]_{\tfrac{9}{11}}(z)
&= z^{-\frac{101}{33}}\left(
1
- \tfrac{239}{33}z
+ \tfrac{428551}{17787}z^{2}
- \tfrac{171087569}{5282739}z^{3}
+ \cdots
\right),
\\
[\tfrac{60}{11},\tfrac{21}{11},\tfrac{38}{11},\tfrac{9}{11}]_{\tfrac{38}{11}}(z)
&= z^{-\frac{14}{33}}\left(
1
- \tfrac{3925}{2508}z
+ \tfrac{965227}{1406988}z^{2}
- \tfrac{3505721843}{8636092344}z^{3}
+ \cdots
\right),
\\
[\tfrac{60}{11},\tfrac{21}{11},\tfrac{38}{11},\tfrac{9}{11}]_{\tfrac{87}{11}}(z)
&= z^{\frac{133}{33}}\left(
1
+ \tfrac{47}{33}z
+ \tfrac{722048}{386595}z^{2}
+ \tfrac{3618078524}{1569189105}z^{3}
+ \cdots
\right),
\end{aligned}
\end{align*}\vspace{-1em}
\begin{align*}
\begin{aligned}
[\tfrac{21}{11},\tfrac{9}{11},\tfrac{38}{11},\tfrac{60}{11}]_{\tfrac{2}{11}}(z)
&= z^{-\frac{122}{33}}\left(
1
+ \tfrac{137}{33}z
- \tfrac{2215}{1089}z^{2}
+ \tfrac{10191745}{754677}z^{3}
+ \cdots
\right),
\\
[\tfrac{21}{11},\tfrac{9}{11},\tfrac{38}{11},\tfrac{60}{11}]_{\tfrac{21}{11}}(z)
&= z^{-\frac{65}{33}}\left(
1
- \tfrac{191}{462}z
+ \tfrac{8078869}{10626462}z^{2}
+ \tfrac{2409984575}{1052019738}z^{3}
+ \cdots
\right),
\\
[\tfrac{21}{11},\tfrac{9}{11},\tfrac{38}{11},\tfrac{60}{11}]_{\tfrac{60}{11}}(z)
&= z^{\frac{52}{33}}\left(
1
+ \tfrac{163}{165}z
+ \tfrac{885881}{664290}z^{2}
+ \tfrac{110917549}{65764710}z^{3}
+ \cdots
\right),
\end{aligned}
\end{align*}\vspace{-1em}
\begin{align*}
\begin{aligned}
[\tfrac{21}{11},\tfrac{60}{11},\tfrac{38}{11},\tfrac{9}{11}]_{\tfrac{9}{11}}(z)
&= z^{-\frac{101}{33}}\left(
1
+ \tfrac{46}{11}z
+ \tfrac{657571}{53361}z^{2}
+ \tfrac{72012964}{5282739}z^{3}
+ \cdots
\right),
\\
[\tfrac{21}{11},\tfrac{60}{11},\tfrac{38}{11},\tfrac{9}{11}]_{\tfrac{38}{11}}(z)
&= z^{-\frac{14}{33}}\left(
1
+ \tfrac{2861}{2508}z
+ \tfrac{1030583}{703494}z^{2}
+ \tfrac{18418358023}{8636092344}z^{3}
+ \cdots
\right),
\\
[\tfrac{21}{11},\tfrac{60}{11},\tfrac{38}{11},\tfrac{9}{11}]_{\tfrac{87}{11}}(z)
&= z^{\frac{133}{33}}\left(
1
+ \tfrac{86}{33}z
+ \tfrac{1871183}{386595}z^{2}
+ \tfrac{12132250892}{1569189105}z^{3}
+ \cdots
\right),
\end{aligned}
\end{align*}\vspace{-1em}
\begin{align*}
\begin{aligned}
[\tfrac{60}{11},\tfrac{9}{11},\tfrac{38}{11},\tfrac{21}{11}]_{\tfrac{21}{11}}(z)
&= z^{-\frac{65}{33}}\left(
1
- \tfrac{661}{231}z
+ \tfrac{161845}{129591}z^{2}
+ \tfrac{58649795}{12829509}z^{3}
+ \cdots
\right),
\\
[\tfrac{60}{11},\tfrac{9}{11},\tfrac{38}{11},\tfrac{21}{11}]_{\tfrac{60}{11}}(z)
&= z^{\frac{52}{33}}\left(
1
+ \tfrac{173}{1320}z
+ \tfrac{26118527}{98314920}z^{2}
+ \tfrac{221986579}{973317708}z^{3}
+ \cdots
\right),
\\
[\tfrac{60}{11},\tfrac{9}{11},\tfrac{38}{11},\tfrac{21}{11}]_{\tfrac{119}{11}}(z)
&= z^{\frac{229}{33}}\left(
1
+ \tfrac{725}{231}z
+ \tfrac{148160}{22869}z^{2}
+ \cdots
\right).
\end{aligned}
\end{align*}
The F-matrices are
\begin{align*}
\begin{gathered}
F_{st}=\small\begin{pmatrix}
1.02301 & -6.30212 & 0.368113 \\
-0.239915 & 0.306438 & 0.0449871 \\
0.153128 & 0.783745 & 0.0205407
\end{pmatrix},\ 
F_{s't'}=\small\begin{pmatrix}
0.549244 & 5.69129 & 1.54463 \\
0.0671232 & -0.276737 & -0.362245 \\
0.0306479 & -0.707781 & 0.231206
\end{pmatrix},\\
F_{u'u}=\small\begin{pmatrix}
0.845234 & 7.27871 & 0.511725 \\
0.188404 & 0.515113 & -0.0353282 \\
0.464713 & -1.70152 & 0.0370501
\end{pmatrix}.
\end{gathered}
\end{align*}
The unitary F-matrices are
\begin{align*}
\begin{gathered}
F_{st} = {\sqrt{\gamma_{11}}}\small\begin{pmatrix}
1 & \sqrt{\gamma_{12}} & -\sqrt{\gamma_{13}} \\
-\sqrt{\gamma_{14}}  & \sqrt{\gamma_{15}}  & \sqrt{\gamma_{16}}  \\
\sqrt{\gamma_{17}}  & \sqrt{\gamma_{18}}  & \sqrt{\gamma_{19}} 
\end{pmatrix}, \quad
F_{s't'} = {\sqrt{\gamma_{11}} }
\small\begin{pmatrix}
\sqrt{\gamma_{13}} & \sqrt{\gamma_{12}}  & 1 \\
\sqrt{\gamma_{16}}  & -\sqrt{\gamma_{15}}  & -\sqrt{\gamma_{14}} \\
\sqrt{\gamma_{19}}  & -\sqrt{\gamma_{18}}  & \sqrt{\gamma_{17}}
\end{pmatrix}, \\
F_{u'u} = {\sqrt{\gamma_{21}} }
\small\begin{pmatrix}
1 & \sqrt{\gamma_{22}}  & \sqrt{\gamma_{23}}  \\
\sqrt{\gamma_{24}} & \sqrt{\gamma_{25}} & -\sqrt{\gamma_{26}} \\
\sqrt{\gamma_{27}} & -\sqrt{\gamma_{24}} & 1
\end{pmatrix}.
\end{gathered}
\end{align*}
Here,
\begin{align*}
\gamma_{11}&=-2+4\cos\tfrac{\pi}{11}-2\cos\tfrac{2\pi}{11},\quad 
\gamma_{12} = 2\cos\tfrac{2\pi}{11} + 2\cos\tfrac{4\pi}{11}, \quad
\gamma_{13} = 1 + 2\cos\tfrac{\pi}{11}, \\
\gamma_{14} &= 2 - 2\cos\tfrac{\pi}{11} - 2\cos\tfrac{6\pi}{11}, \quad
\gamma_{15} = -1 + 4\cos\tfrac{\pi}{11} + 2\cos\tfrac{4\pi}{11} + 2\cos\tfrac{6\pi}{11}, \\
\gamma_{16} &= 1 + 2\cos\tfrac{2\pi}{11}, \quad
\gamma_{17} = 2 - 2\cos\tfrac{\pi}{11} + 2\cos\tfrac{2\pi}{11} - 2\cos\tfrac{6\pi}{11}, \\
\gamma_{18} &= 4\cos\tfrac{\pi}{11} + 2\cos\tfrac{6\pi}{11}, \quad
\gamma_{19} = 2\cos\tfrac{4\pi}{11},
\end{align*}~\vspace{-2em}
\begin{align*}
\gamma_{21}&=-5 + 10\cos\tfrac{\pi}{11} - 6\cos\tfrac{2\pi}{11} + 4\cos\tfrac{4\pi}{11} +  8\cos\tfrac{6\pi}{11}, \\
\gamma_{22} &= 1 + 4\cos\tfrac{2\pi}{11} + 2\cos\tfrac{4\pi}{11} + 2\cos\tfrac{6\pi}{11}, \quad
\gamma_{23} = 3 + 2\cos\tfrac{\pi}{11} + 2\cos\tfrac{2\pi}{11} + 4\cos\tfrac{4\pi}{11}, \\
\gamma_{24} &= 2 + 4\cos\tfrac{2\pi}{11} + 2\cos\tfrac{4\pi}{11}, \quad
\gamma_{25} = 2 + 2\cos\tfrac{\pi}{11} - 2\cos\tfrac{2\pi}{11}  + 2\cos\tfrac{4\pi}{11}, \\
\gamma_{26} &= 1 + 4\cos\tfrac{2\pi}{11} + 2\cos\tfrac{4\pi}{11} + 2\cos\tfrac{6\pi}{11}, \quad
\gamma_{27} = 2 + 2\cos\tfrac{\pi}{11} + 2\cos\tfrac{2\pi}{11} + 4\cos\tfrac{4\pi}{11} +  2\cos\tfrac{6\pi}{11}.
\end{align*}
The values $\gamma_i\in \mathbb{Z}[\zeta_{22}]$ and the squares of the 
entries of the $\bar{F}$-matrix lie in $\mathbb{Q}(\zeta_{22})\subset 
\mathbb{Q}(\zeta_{66})$.

\paragraph{Order-4$^{\dagger}$:} We consider the non-unitary minimal 
model $\mathcal{M}(10,7)$ with central charge $c=8/35$ and 27 primaries. 
We will analyse the correlator 
$\langle\phi_{\frac98}\phi_{\frac{11}{5}}\phi_{\frac{11}{56}}\phi_{\frac{27}{35}}\rangle$. 
There are 24 distinct conformal blocks based on 6 channels each 
containing 4 intermediate primaries:
\begin{align*}
\begin{aligned}
[\tfrac{9}{8},\tfrac{11}{5},\tfrac{11}{56},\tfrac{27}{35}]_{\tfrac{1}{40}}(z)
&= z^{-\frac{1181}{840}}\left(
1
- \tfrac{2195}{168}z
+ \tfrac{1755701}{282240}z^{2}
- \tfrac{35450232257}{13513651200}z^{3}
+ \cdots
\right),
\\
[\tfrac{9}{8},\tfrac{11}{5},\tfrac{11}{56},\tfrac{27}{35}]_{\tfrac{9}{8}}(z)
&= z^{-\frac{257}{840}}\left(
1
- \tfrac{5389}{12600}z
+ \tfrac{477311941}{3281040000}z^{2}
+ \tfrac{13261790435927}{62777232000000}z^{3}
+ \cdots
\right),
\\
[\tfrac{9}{8},\tfrac{11}{5},\tfrac{11}{56},\tfrac{27}{35}]_{\tfrac{29}{8}}(z)
&= z^{\frac{1843}{840}}\left(
1
+ \tfrac{123233}{121800}z
+ \tfrac{20044711847}{19439280000}z^{2}
+ \tfrac{2832399189262273}{2694284208000000}z^{3}
+ ...
\right),
\\
[\tfrac{9}{8},\tfrac{11}{5},\tfrac{11}{56},\tfrac{27}{35}]_{\tfrac{301}{40}}(z)
&= z^{\frac{5119}{840}}\left(
1
+ \tfrac{505}{168}z
+ \tfrac{97025359}{16652160}z^{2}
+ \tfrac{388964446883}{41963443200}z^{3}
+ \cdots
\right).
\end{aligned}
\end{align*}\vspace{-1em}
\begin{align*}
\begin{aligned}
[\tfrac{9}{8},\tfrac{27}{35},\tfrac{11}{56},\tfrac{11}{5}]_{\tfrac{27}{280}}(z)
&= z^{-\frac{1121}{840}}\left(
1
+ \tfrac{505}{168}z
+ \tfrac{73381319}{5362560}z^{2}
+ \tfrac{1766625775151}{391895884800}z^{3}
+ \cdots
\right),
\\
[\tfrac{9}{8},\tfrac{27}{35},\tfrac{11}{56},\tfrac{11}{5}]_{\tfrac{11}{56}}(z)
&= z^{-\frac{1037}{840}}\left(
1
+ \tfrac{4981}{4200}z
+ \tfrac{247635649}{35280000}z^{2}
+ \tfrac{30458666456347}{13780368000000}z^{3}
+ \cdots
\right),
\\
[\tfrac{9}{8},\tfrac{27}{35},\tfrac{11}{56},\tfrac{11}{5}]_{\tfrac{95}{56}}(z)
&= z^{\frac{223}{840}}\left(
1
+ \tfrac{136271}{399000}z
+ \tfrac{92528223}{372400000}z^{2}
+ \tfrac{5136960507767}{42230160000000}z^{3}
+ \cdots
\right),
\\
[\tfrac{9}{8},\tfrac{27}{35},\tfrac{11}{56},\tfrac{11}{5}]_{\tfrac{1287}{280}}(z)
&= z^{\frac{2659}{840}}\left(
1
+ \tfrac{3625}{2184}z
+ \tfrac{7960913}{3669120}z^{2}
+ \tfrac{7070834529223}{2727623808000}z^{3}
+ \cdots
\right).
\end{aligned}
\end{align*}\vspace{-1em}
\begin{align*}
\begin{aligned}
[\tfrac{27}{35},\tfrac{11}{5},\tfrac{11}{56},\tfrac{9}{8}]_{\tfrac{4}{7}}(z)
&= z^{-\frac{361}{420}}\left(
1
- \tfrac{167}{105}z
+ \tfrac{431131}{1411200}z^{2}
+ \tfrac{1164097057}{1778112000}z^{3}
+ \cdots
\right),
\\
[\tfrac{27}{35},\tfrac{11}{5},\tfrac{11}{56},\tfrac{9}{8}]_{-\tfrac{1}{35}}(z)
&= z^{-\frac{613}{420}}\left(
1
+ \tfrac{18887}{840}z
- \tfrac{23887499}{1411200}z^{2}
+ \tfrac{6407486069}{7823692800}z^{3}
+ \cdots
\right),
\\
[\tfrac{27}{35},\tfrac{11}{5},\tfrac{11}{56},\tfrac{9}{8}]_{\tfrac{27}{35}}(z)
&= z^{-\frac{277}{420}}\left(
1
- \tfrac{8993}{7560}z
+ \tfrac{1811581}{12700800}z^{2}
+ \tfrac{61572686659}{202704768000}z^{3}
+ \cdots
\right),
\\
[\tfrac{27}{35},\tfrac{11}{5},\tfrac{11}{56},\tfrac{9}{8}]_{\tfrac{104}{35}}(z)
&= z^{\frac{647}{420}}\left(
1
+ \tfrac{919}{1680}z
+ \tfrac{836447}{1764000}z^{2}
+ \tfrac{201538385213}{462309120000}z^{3}
+ \cdots
\right).
\end{aligned}
\end{align*}\vspace{-1em}
\begin{align*}
\begin{aligned}
[\tfrac{11}{5},\tfrac{9}{8},\tfrac{11}{56},\tfrac{27}{35}]_{\tfrac{1}{40}}(z)
&= z^{-\frac{1181}{840}}\left(
1
+ \tfrac{4897}{420}z
+ \tfrac{424069}{352800}z^{2}
+ \tfrac{40538645047}{8446032000}z^{3}
+ \cdots
\right),
\\
[\tfrac{11}{5},\tfrac{9}{8},\tfrac{11}{56},\tfrac{27}{35}]_{\tfrac{9}{8}}(z)
&= z^{-\frac{257}{840}}\left(
1
+ \tfrac{767}{6300}z
+ \tfrac{275727209}{820260000}z^{2}
+ \tfrac{16008471093359}{70624386000000}z^{3}
+ \cdots
\right),
\\
[\tfrac{11}{5},\tfrac{9}{8},\tfrac{11}{56},\tfrac{27}{35}]_{\tfrac{29}{8}}(z)
&= z^{\frac{1843}{840}}\left(
1
+ \tfrac{72001}{60900}z
+ \tfrac{6334604603}{4859820000}z^{2}
+ \tfrac{469877607723749}{336785526000000}z^{3}
+ \cdots
\right),
\\
[\tfrac{11}{5},\tfrac{9}{8},\tfrac{11}{56},\tfrac{27}{35}]_{\tfrac{301}{40}}(z)
&= z^{\frac{5119}{840}}\left(
1
+ \tfrac{1297}{420}z
+ \tfrac{18192353}{2973600}z^{2}
+ \tfrac{259936065167}{26227152000}z^{3}
+ \cdots
\right).
\end{aligned}
\end{align*}\vspace{-1em}
\begin{align*}
\begin{aligned}
[\tfrac{11}{5},\tfrac{27}{35},\tfrac{11}{56},\tfrac{9}{8}]_{\tfrac{4}{7}}(z)
&= z^{-\frac{361}{420}}\left(
1
+ \tfrac{307}{420}z
+ \tfrac{661231}{1411200}z^{2}
- \tfrac{179415259}{889056000}z^{3}
+ \cdots
\right),
\\
[\tfrac{11}{5},\tfrac{27}{35},\tfrac{11}{56},\tfrac{9}{8}]_{-\tfrac{1}{35}}(z)
&= z^{-\frac{613}{420}}\left(
1
- \tfrac{20113}{840}z
- \tfrac{8833499}{1411200}z^{2}
- \tfrac{55666419331}{7823692800}z^{3}
+ \cdots
\right),
\\
[\tfrac{11}{5},\tfrac{27}{35},\tfrac{11}{56},\tfrac{9}{8}]_{\tfrac{27}{35}}(z)
&= z^{-\frac{277}{420}}\left(
1
+ \tfrac{4007}{7560}z
+ \tfrac{5529581}{12700800}z^{2}
+ \tfrac{22040349659}{202704768000}z^{3}
+ \cdots
\right),
\\
[\tfrac{11}{5},\tfrac{27}{35},\tfrac{11}{56},\tfrac{9}{8}]_{\tfrac{104}{35}}(z)
&= z^{\frac{647}{420}}\left(
1
+ \tfrac{1669}{1680}z
+ \tfrac{3673519}{3528000}z^{2}
+ \tfrac{504874270463}{462309120000}z^{3}
+ \cdots
\right).
\end{aligned}
\end{align*}\vspace{-1em}
\begin{align*}
\begin{aligned}
[\tfrac{27}{35},\tfrac{9}{8},\tfrac{11}{56},\tfrac{11}{5}]_{\tfrac{27}{280}}(z)
&= z^{-\frac{1121}{840}}\left(
1
- \tfrac{1823}{420}z
+ \tfrac{99963391}{6703200}z^{2}
+ \tfrac{1220410369883}{244934928000}z^{3}
+ \cdots
\right),
\\
[\tfrac{27}{35},\tfrac{9}{8},\tfrac{11}{56},\tfrac{11}{5}]_{\tfrac{11}{56}}(z)
&= z^{-\frac{1037}{840}}\left(
1
- \tfrac{5083}{2100}z
+ \tfrac{65638861}{8820000}z^{2}
+ \tfrac{5694890084951}{1722546000000}z^{3}
+ \cdots
\right),
\\
[\tfrac{27}{35},\tfrac{9}{8},\tfrac{11}{56},\tfrac{11}{5}]_{\tfrac{95}{56}}(z)
&= z^{\frac{223}{840}}\left(
1
- \tfrac{15173}{199500}z
- \tfrac{1467093}{93100000}z^{2}
+ \tfrac{414535317031}{5278770000000}z^{3}
+ \cdots
\right),
\\
[\tfrac{27}{35},\tfrac{9}{8},\tfrac{11}{56},\tfrac{11}{5}]_{\tfrac{1287}{280}}(z)
&= z^{\frac{2659}{840}}\left(
1
+ \tfrac{8221}{5460}z
+ \tfrac{8478937}{4586400}z^{2}
+ \tfrac{143905608607}{68190595200}z^{3}
+ \cdots
\right).
\end{aligned}
\end{align*}
The exponents are easily read off, and one finds an integer difference: 
$\frac{647}{420}-\frac{-613}{420}=3$. Consequently, the Frobenius 
solution is underdetermined, and we require the conformal blocks to fix 
$a_{3}$. The computational details are straightforward and we obtain the 
orthogonal F-matrices:
\begin{align*}
\begin{gathered}
\bar{F}_{st} = {\sqrt{\gamma_1}} \small\begin{pmatrix}
1 & -\sqrt{\gamma_3} & \sqrt{\gamma_4} & \sqrt{\gamma_5} \\
-\sqrt{\gamma_3} & \sqrt{\gamma_6} & 0 & -\sqrt{\gamma_4} \\
\sqrt{\gamma_4} & 0 & -\sqrt{\gamma_6} & -\sqrt{\gamma_3} \\
\sqrt{\gamma_5} & -\sqrt{\gamma_4 }& -\sqrt{\gamma_3} & -1
\end{pmatrix},\quad
\bar{F}_{s't'}= {\sqrt{\gamma_2}}\small \begin{pmatrix}
-1 & -\sqrt{\gamma_9} & \sqrt{\gamma_7} & \sqrt{\gamma_8} \\
-\sqrt{\gamma_7} & -\sqrt{\gamma_8} & 1 & \sqrt{\gamma_9} \\
-\sqrt{\gamma_7} & \sqrt{\gamma_8} & 1 & -\sqrt{\gamma_9} \\
-1 & \sqrt{\gamma_9} & \sqrt{\gamma_7} & -\sqrt{\gamma_8}
\end{pmatrix},\\
\bar{F}_{u'u}= {\sqrt{\gamma_2}}\small \begin{pmatrix}
-1 & \sqrt{\gamma_7} & -\sqrt{\gamma_7}  & 1 \\
-\sqrt{\gamma_9 }&  \sqrt{\gamma_8}  & \sqrt{\gamma_8}  & -\sqrt{\gamma_9}  \\
-\sqrt{\gamma_6}  & 1 & -1 & \sqrt{\gamma_7}  \\
-\sqrt{\gamma_8}  & \sqrt{\gamma_9}  & \sqrt{\gamma_9}  & -\sqrt{\gamma_8} 
\end{pmatrix},
\end{gathered}
\end{align*}
where
\begin{align*}
\gamma_1 &= \tfrac14\left(5+6\cos\tfrac{42\pi}{105}\right), \quad
\gamma_2 =1+\cos\tfrac{42\pi}{105},
\\
\gamma_3 &= 1 - 6\cos\tfrac{42\pi}{105},\quad
\gamma_4 = 3 - 10\cos\tfrac{42\pi}{105},\quad
\gamma_5 = 3 - 8\cos\tfrac{42\pi}{105}, \\
\gamma_6 &= 4 - 8\cos\tfrac{42\pi}{105}, \quad
\gamma_7 = 2\cos\tfrac{42\pi}{105}, \quad
\gamma_8 = \tfrac{1}{2}, \quad
\gamma_9 = \tfrac12\left(1 - 4\cos\tfrac{42\pi}{105}\right).
\end{align*}
Observe that for $i\in [3,8]$, one has $\gamma_i \in 
\mathbb{Z}[\zeta_{105}]$ but $\gamma_1,\gamma_2,\gamma_9\in 
\mathbb{Q}(\zeta_{105})$. Thus, the squares of the entries of the 
$\bar{F}$-matrix lie in 
$\mathbb{Q}(\zeta_{105})\subset\mathbb{Q}(\zeta_{1680})$.

\section{Symmetries that extend Virasoro algebra}
\label{sec:extended}

The holomorphic modular bootstrap produces RCFTs and tenable characters 
associated with symmetries that extend the Virasoro symmetry, i.e. 
characters not of Virasoro algebra but of some larger symmetry. These 
could be simple current extensions, affine algebras, W-algebras or 
combinations thereof. In this section, we discuss how the conformal 
block for correlators in these extended theories can be written down 
using the Virasoro conformal blocks derived in 
Section~\ref{sec:blockcompute}. We will then illustrate the result with 
some simple examples. The basic idea is that the extended characters are 
reducible when considered as characters of the Virasoro algebra.

\subsection{The direct computation revisited}

When an RCFT has an extended symmetry, then the primaries are 
irreducible under the extended symmetry. Examples include RCFTs 
associated with affine Kac-Moody Lie algebras, W${_n}$ algebra, etc. The 
irreducible modules in this case are reducible as Virasoro modules. 
Non-diagonal partition functions of minimal models can also be 
understood in this fashion. Let $M_h$ be an extended symmetry module 
with conformal weight $h$ and central charge $c$. We write

\begin{equation}
    M_h \cong \oplus_\alpha m_\alpha\, V^c_{h_\alpha},\qquad 
    h_\alpha=h+s_\alpha,\quad s_\alpha\in \mathbb{Z}_{\geq0},
\end{equation}
where $V^c_{h_\alpha}$ is an irreducible Virasoro Verma module with 
central charge $c$ and weight $h_\alpha$. The sum over alpha, which need 
not be finite, runs over all the Virasoro irreps that are contained in 
the module $M_h$ and $m_\alpha\in \mathbb{Z}_{\geq0}$ are the 
multiplicities. The corresponding character is given by
\[
\chi_h(q) := \text{Tr}_{M_h} \left(q^{L_0-\frac{c}{24}}\right) =
\sum_\alpha m_\alpha\, \chi_{h+s_\alpha}^{\rm Vir}(q) 
\]
where the sum is over Virasoro characters. We assume that the torus 
partition function is diagonal in the extended characters.

We will now incorporate this into the computation of the four-point 
function done in Section~\ref{sec:blockcompute}. First, secondaries in 
the module now carry two labels, the Greek letter $\alpha$ to indicate 
the Virasoro primary and the partition $\lambda$ to label the secondary 
under the Virasoro primary. The three-point function between two 
primaries and one secondary now takes the form
\begin{equation}\label{3ptaext} \tag{\ref*{3pta}$'$}
 \langle \phi_{p,\alpha}^\lambda| \phi_k(z) |\phi_l\rangle   = {c_{kl}}^p\ z^{h_p+|\lambda|+s_\alpha-h_k-h_l} \sum_{\lambda'} \mathcal{M}_{\lambda\lambda'}^{(p,\alpha)}\ \beta_{kl}^{(p,\alpha),\lambda'}.
 \end{equation}
where we define the Shapovalov form over the $\alpha$-th primary by
 \[
 \mathcal{M}_{\lambda\lambda'}^{(p,\alpha)}= \langle\phi_{(p,\alpha)}^\lambda|\phi_{(p,\alpha)}^{\lambda'}\rangle.
 \]
Another computation of the three-point function computed 
in~\eqref{3ptaext} is as follows.
\begin{equation}\label{3ptbext}\tag{\ref*{3ptb}$'$}
\langle \phi^\lambda_{(p,\alpha)}| \phi_k(z) |\phi_l\rangle =
\langle \phi_{(p,\alpha)}| \mathcal{L}_\lambda\phi_k(z) |\phi_l\rangle  
= {c_{kl}}^p\ z^{h_p+s_\alpha+|\lambda|-h_k-h_l}\ \alpha_{kl}^{(p,\alpha),\lambda}
\end{equation}
where
\[
 \alpha_{kl}^{(p,\alpha),\lambda}  =(z)^{-h_p-s_\alpha+h_k+h_l-|\lambda|} \hat{L}_{\ell_s}\cdots \hat{L}_{\ell_1}\cdot (z)^{h_p+s_\alpha-h_k-h_l},
\]
and $\hat{L}_\ell =z^{\ell+1}\partial_z + h_k\ell+1)z^\ell$. An explicit formula is then given by
\begin{equation}\label{alphaformulaext}\tag{\ref*{alphaformula}$'$}
\alpha_{kl}^{(p,\alpha),(\ell_1,\ldots,\ell_s)}= \prod_{x=1}^s \Big[(\ell_x+1)h_k +(h_p+s_\alpha -h_k-h_l)+N_x\Big]\ ,
\end{equation}
with $N_x =\sum_{i=1}^{x-1}\ell_i$. Comparing~\eqref{3ptaext} and 
\eqref{3ptbext}, for any two partitions $\lambda$ and $\lambda'$ with 
$|\lambda|=|\lambda'|$. we obtain the relation
 \begin{equation}\label{alphabetaext}\tag{\ref*{alphabeta}$'$}
\alpha_{kl}^{(p,\alpha),\lambda} := \sum_{\lambda'} \mathcal{M}_{\lambda\lambda'}^{(p,\alpha)}\ \beta_{kl}^{(p,\alpha),\lambda'}.
\end{equation}

The four-point function is thus
\begin{equation} \label{Fblockdefext}
\langle \phi_i| \phi_j(1) \phi_k(z) |\phi_l\rangle =
\sum_{p} {c_{ij}}^p c_{kl}{}^p \left|\sum_\alpha  g_\alpha\, \langle ijkl\rangle_{(p,\alpha)}(z)\right|^2\ ,
\end{equation}
where $\langle ijkl\rangle_{(p,\alpha)}(z)$ is the chiral conformal 
block associated with the primary labeled $\alpha$ under the extended 
primary labeled $p$ with
\begin{align}\label{feqnext}
\langle ijkl\rangle_{(p,\alpha)}(z):= z^{h_p+s_\alpha -h_k-h_l}\sum_{\lambda} \left(\beta_{ji}^{(p,\alpha),\lambda}\right)^* \cdot \alpha_{kl}^{(p,\alpha),\lambda}\ z^{|\lambda|},
\end{align}
and the gluing coefficients $g_\alpha$ are implicitly defined above. 
These remain unfixed for the moment like the three-point structure 
constants.

Another generalization is when the reducible module $M_h$ arises from a 
tensor product of two RCFTs.
\begin{equation}
    M_h = \oplus_\alpha \left(V^{c'}_{h'_\alpha}\otimes V^{c''}_{h''_\alpha} \right)\ ,
\end{equation}
with $c=c'+c''$ and $h=h'_\alpha+h''_\alpha$. A straightforward 
computation (not given here) shows that the block takes the form
\[
\langle ijkl\rangle_{(p,\alpha)}(z):= z^{h_p+s'_\alpha+s''_\alpha -h_k-h_l}\sum_{\boldsymbol\lambda} (\beta_{ji}^{(p,\alpha),\boldsymbol\lambda})^* \cdot \alpha_{kl}^{(p,\alpha),\boldsymbol\lambda}\ z^{|\boldsymbol\lambda|}\ ,
\]
with the two partitions $\boldsymbol{\lambda}=(\lambda',\lambda'')$ 
representing secondaries in the sub-theories, $|\boldsymbol{\lambda}| 
=|\lambda'|+|\lambda''|$ and the Shapovalov matrices and $\alpha$ and 
$\beta$ taken as `products' of the two sub-theories.

\subsection{An example: The three-state Potts model}

The three-state Potts model is a finite extension of the minimal model 
${\cal M}(6,5)$ with central charge $c=4/5$, and is described by a 
non-diagonal modular invariant with respect to Virasoro algebra. To 
write down the modular invariant, consider the conformal weights of the 
minimal model:
\begin{align}
    h_{r,s}=\frac{(6r-5s)^2-1}{120}, \qquad r=1,\hdots, 4,\quad s=1,\hdots,5,
\end{align}
with the identification $(r,s)\sim(5-r,6-s)$, i.e. there are 10 Virasoro primaries (corresponding to $r=1,2,\ s=1,\hdots,5$):
$$
h\in\left\{0,\frac18,\frac23,\frac{13}{8},3,\frac25,\frac{1}{40},\frac{1}{15},\frac{21}{40},\frac75\right\}.
$$
Now, consider the block characters:
\begin{align}
\begin{aligned}
      \bar\chi_{r,1}&=\chi_{r,1}+\chi_{r,5},\\
      \bar\chi_{r,3}&=\chi_{r,3},
\end{aligned}
\end{align}
with $r=1,2$. Then, it turns out that 
\begin{equation}\label{PottsZ}
Z_{\rm Potts}=\sum_{r=1,2}\left(|\bar\chi_r,1|^2+2|\bar \chi_{r,3}|^2\right)=\sum_{r=1,2}\left(\big|\chi_{r,1}+\chi_{r,5}\big|^{2}
+2\big|\chi_{r,3}\big|^{2}\right)
\end{equation}  
is modular invariant. Note that $|\chi_{1,1}+\chi_{1,5}|^2$ contains the 
cross-term $\chi_{1,5}\bar \chi_{1,1}$, i.e. an operator with
$$
(h,\bar h)=(3,0).
$$
This is a purely holomorphic field of spin-3, i.e. a conserved chiral 
current beyond the spin-2 stress tensor. Thus, the algebra underlying 
Potts model is W$_3$. The field content is:
\begin{align*}
\begin{tabular}{c|cc}
  W$_3$ module & Virasoro content  & Weight \\\hline
$[\mathbf{1}]$     & $\phi_{0}\oplus\phi_{3}$ & $0$\\ 
  $[\varepsilon]$   & $\phi_{\frac25}\oplus\phi_{\frac75}$  & $\frac25$\\ 
 $[\sigma]$, $[\sigma^\dagger]$    & $\phi_{\frac{1}{15}}$ each & $\frac{1}{15}$ \\ 
$[Z]$, $[Z^\dagger]$     & $\phi_{\frac23}$ each & $\frac23$
\end{tabular}    
\end{align*}
It is important to note that the `long' primaries (that involve more 
than one Virasoro primaries) take the lowest of the weights of the 
contributing Virasoro primaries. This is because the higher weights show 
up as W$_3$-descendants in the module. Further, the model exhibits 
$\mathbb{Z}_3$ permutation symmetry, under which we have charges $+1$ 
for $\sigma,Z$; $-1$ for $\sigma^\dagger,Z^\dagger$; 0 for 
$\mathbf{1},\varepsilon$. Thus, one can write down the fusion rules 
consistent with $\mathbb{Z}_3$-charge conservation:
\begin{align}\label{eq:pottsfusion}
    \sigma\times \sigma^\dagger=\mathbf{1}+\varepsilon,
    \quad
    \sigma\times \sigma=\sigma^\dagger+Z^\dagger,
    \quad
    \varepsilon\times \varepsilon=\mathbf{1}+\varepsilon
\end{align}
It is easy to check that these rules are also consistent with Virasoro 
fusion rules, except that Virasoro fusion knows nothing about 
$\mathbb{Z}_3$, so it returns the union of all charge sectors. For 
example, Virasoro fusion gives us
\begin{align}
    \phi_{\frac{1}{15}}\times \phi_{\frac{1}{15}}=\phi_{0}+\phi_{\frac23}+\phi_{3}+\phi_{\frac25}+\phi_{\frac{1}{15}}+\phi_{\frac75},
\end{align}
which gives two of the three Potts fusion rules in~\eqref{eq:pottsfusion}. 

We are ready to write down the conformal blocks for correlators in Potts 
model using the Virasoro core blocks:
\begin{align}
    {\cal F}_P(z)=\sum_{p\in P} g_p V_p(z),
\end{align}
where ${\cal F}_{[P]}$ is the Potts block for intermediate primary 
$[P]$, $V_p(z)$ is an ordinary Virasoro core block, and $g_p$ are the 
gluing coefficients (normalized to 1 for the lowest-weight member).

We consider the correlator 
$\langle\sigma\sigma^\dagger\sigma\sigma^\dagger\rangle$. All external 
operators have $h_\sigma=1/15$. For the $s$-channel ($z\to 0$), we get
\begin{align}
    \begin{aligned}
    [\sigma\sigma^\dagger\sigma\sigma^\dagger]_\mathbf{1}&=[\tfrac{1}{15}\tfrac{1}{15}\tfrac{1}{15}\tfrac{1}{15}]_0 + g_{3}\, [\tfrac{1}{15}\tfrac{1}{15}\tfrac{1}{15}\tfrac{1}{15}]_3,\\
    [\sigma\sigma^\dagger\sigma\sigma^\dagger]_\mathbf{\varepsilon}&=[\tfrac{1}{15}\tfrac{1}{15}\tfrac{1}{15}\tfrac{1}{15}]_\frac{2}{5} + g_{\frac75}\, [\tfrac{1}{15}\tfrac{1}{15}\tfrac{1}{15}\tfrac{1}{15}]_\frac{7}{5},
    \end{aligned}\label{eq:pottsblock}
\end{align}
where $g_3,g_{\frac75}$ are yet undetermined.

The null behaviour of the Virasoro primary $\phi_\frac{1}{15}$ is 
irrelevant here since Potts exhibits W$_3$ symmetry, instead of Virasoro 
symmetry, as we have shown. Hence, we cannot rely on the BPZ equation. 
To obtain the Fuchsian ODE (with three singularities) that the Potts 
conformal blocks satisfy, we can recast the Fuchs 
relation~\eqref{eq:fuchs} as
\begin{align}
    \ell=\frac{n(n-1)}{2}-\left[\sum_p h_p+\sum_q h_q+\sum_r h_r-n\,h\right],
\end{align}
where $n$ is the number of exchanged primaries. Here, $\ell\neq 0$ 
suggests that the Fuchs relation is not satisfied, and that there are 
additional singularities, or alternatively one needs to consider 
spurious solutions (refer Section~\ref{sec:special}). For 
$\langle\sigma\sigma^\dagger\sigma\sigma^\dagger\rangle$, with $n=2$, it 
is easy to check that one gets $\ell=0$ and the blocks 
in~\eqref{eq:pottsblock} are solutions of an order-2 BPZ-like Fuchsian 
ODE. Since there are no accessory parameters at order-2, the ODE is 
determined solely by the local exponents:
\begin{equation}
    \left[\dfrac{d^2}{dz^2} +\frac{(14z-7)}{9z(z-1)} \dfrac{d}{dz} 
    + \frac{-26(z^2-z)-56}{2025\,z^2(z-1)^2}\right][\sigma\sigma^\dagger\sigma\sigma^\dagger](z) = 0.
\end{equation}
The ODE in turn determines for us the gluing coefficients which happen 
to be ratios of the Virasoro OPE coefficients:
\begin{align}
    g_3=\frac{1}{2106}=\frac{c_{\frac{1}{15}\frac{1}{15}} {}^{3}}{c_{\frac{1}{15}\frac{1}{15}} {}^{0}},\qquad g_{\frac{7}{5}}=\frac{1}{21}=\frac{c_{\frac{1}{15}\frac{1}{15}} {}^{\frac75}}{c_{\frac{1}{15}\frac{1}{15}} {}^{\frac25}}.
\end{align}
The knowledge of the ODE now simply leads to the braiding matrices:
\begin{align}
F_{st}={\small\begin{pmatrix}
0.61803 & 0.42938 \\
1.4394 & -0.61803
\end{pmatrix}},\quad
F_{s't'}=F_{u'u}=\small\begin{pmatrix}
0.85876 & 0.20601 \\
1.2361 & -0.47979
\end{pmatrix}.
\end{align}
Note there are two unique ones since 
$\langle\sigma\sigma^\dagger\sigma\sigma^\dagger\rangle$ contains 
pairwise identical operators (see Section \ref{sec:correlatorsymmetry}). 
The unitary versions are
\begin{align}
\bar F_{st}={\small\begin{pmatrix}
\gamma & \sqrt{\gamma} \\
\sqrt{\gamma} & -\gamma
\end{pmatrix}},\quad
\bar F_{s't'}=\bar F_{u'u}=\small\begin{pmatrix}
\sqrt{\gamma} & \gamma \\
\gamma & -\sqrt\gamma
\end{pmatrix},
\end{align}
where $\gamma=2\cos\frac{2\pi}{5}\in \mathbb{Q}(\zeta_5)$.

Let us consider another correlator $\langle 
\varepsilon\varepsilon\varepsilon\varepsilon\rangle$, with 
$h_\varepsilon=\frac25$. The fusion rules need to be analyzed carefully. 
Note the Virasoro fusion:
\begin{align}
    \phi_{\frac25}\times\phi_{\frac25}=\phi_0+\phi_{\frac75}.
\end{align}
Comparing this to W$_3$ fusion in~\eqref{eq:pottsfusion}, only Virasoro 
primary $\phi_0$ contributes for $[\mathbf{1}]$, and only $\phi_\frac75$ 
contributes for $[\varepsilon]$. This means that there is no gluing, and 
each W$_3$ conformal block 
$[\varepsilon\varepsilon\varepsilon\varepsilon]$ will be a single 
Virasoro block:
\begin{align}
\begin{aligned}
[\varepsilon\varepsilon\varepsilon\varepsilon]_\mathbf{1}&=[\tfrac25\tfrac25\tfrac25\tfrac25]_0,\qquad
[\varepsilon\varepsilon\varepsilon\varepsilon]_\mathbf{\varepsilon}&=[\tfrac25\tfrac25\tfrac25\tfrac25]_\frac75.
\end{aligned}
\end{align}
Since the correlator contains all identical operators, all the OPE 
channels give the same conformal blocks. The order-2 ODE turns out to be
\begin{equation}
    \left[\dfrac{d^2}{dz^2} +\frac{(4z-2)}{3z(z-1)} \dfrac{d}{dz} 
    - \frac{104(z^2-z+1)}{225\,z^2(z-1)^2}\right][\varepsilon\varepsilon\varepsilon\varepsilon](z) = 0,
\end{equation}
which matches with the BPZ equation for $\phi_\frac25$. Since the 
correlator contains identical operators, there is only one unique 
F-matrix:
\begin{align}
F_{st}={\small\begin{pmatrix}
0.618033 & 0.736077 \\
0.839632 & -0.618033
\end{pmatrix}}\quad\leftrightarrow\quad
\bar F_{st}={\small\begin{pmatrix}
\gamma & \sqrt{\gamma} \\
\sqrt{\gamma} & -\gamma
\end{pmatrix}},
\end{align}
with $\gamma=2\cos\frac{2\pi}{5}\in \mathbb{Q}(\zeta_5)$.

Thus, the formalism adopted for Virasoro conformal blocks and braiding 
matrices can be carefully generalized to compute blocks and braiding 
matrices in RCFTs with algebras that extend the Virasoro algebra.
Non-diagonal minimal models offer a natural testing ground. Their bulk
OPE coefficients have been studied extensively: the D-series
in~\cite{Runkel:1999dz}, and the E-series
in~\cite{Petkova:1994zs}. Most recently, the authors in~\cite{Lhoste:2026qce} obtained the complete
E-series data. These works take the chiral fusion matrices as known,
via the Coulomb-gas or quantum-group $6j$ description, and solve the
crossing equations for the OPE coefficients. In the construction of
this section, both the braiding matrices and the gluing coefficients
are instead determined by the Fuchsian ODE. The gluing coefficients are
ratios of OPE coefficients, closely related to the relative structure
constants of non-diagonal theories studied by Petkova and
Zuber~\cite{Petkova:1994zs}. We will develop this formalism in detail,
and compare it with these results, in a subsequent paper.

\section{Conclusion and discussion}
\label{sec:conclusion}

In this paper, we have pursued an approach that combines direct 
computation as well as the ODE for four-point functions to compute 
braiding matrices (F-matrices and R-matrices) in RCFTs with Virasoro 
symmetry. We have been able to compute the F-matrices directly as 
connection matrices in the ODE. This is first done numerically. We make 
a change of basis to make the F-matrix unitary and determine products of 
three-point structure constants as a by-product. Using Conjecture 
\ref{conjecture}, we have been able to convert numerical values to exact 
values. The R-matrices are calculated trivially through the conformal 
weight of the exchanged primaries. We also verify that together, these 
braiding matrices satisfy the hexagon relation~\cite{Moore:1988qv}.

The formalism developed in this paper can be utilized to compute the 
four-point function involving the primary fields of an extended algebra. 
We present evidence for it by studying the Potts model, which exhibits 
W$_3$ symmetry, and computing a few of its correlators and braiding 
matrices. Furthermore, we can also study RCFTs exhibiting symmetry 
associated with affine Kac--Moody algebras. In these cases, the 
correlators satisfy the celebrated Knizhnik--Zamolodchikov (KZ) 
equations~\cite{Knizhnik:1984nr}. Using $\mathrm{SL}(2,\mathbb{C})$ 
invariance, these equations reduce to Fuchsian ODEs. We can also compute 
the correlation functions using the direct method with Virasoro Verma 
modules being replaced by Verma modules for the affine Lie algebra and 
work with Shapovalov matrices suitable for these Verma modules. The 
F-matrices in these cases are related to the $6j$-symbols for the 
corresponding quantum group. These are not known explicitly for all 
quantum groups but there are some results available~\cite{Nawata:2013ooa,Gu:2014nva}. Our formalism provides an independent way to 
compute them. We plan to fully develop this formalism for the affine cases
in our subsequent paper~\cite{affine}.

Our ultimate goal is to extend these methods to \textit{tenable} 
solutions~\cite{Govindarajan:2026cgv}, where we will be able to identify 
their MTC class by computing the F-matrix for these solutions. This 
might also help in identifying if they are RCFTs. Not every tenable 
solution needs to correspond to an RCFT. There are infinite families of 
solutions with Wronskian index $\ell\geq 6$ that can be constructed from 
tenable solutions with $\ell<6$ by taking linear combinations with 
quasi-characters obtained through the theory of vector-valued modular 
forms (VVMFs)~\cite{Govindarajan:2025rgh,Govindarajan:2025jlq}. Each 
such family shares the S-matrix of the solution one begins with and 
therefore inherits its fusion rules and is itself tenable. Modular data 
alone thus cannot decide whether a given tenable solution is realized by 
an RCFT, and additional constraints are needed. One approach, suggested 
to us by Gannon, is to construct multi-variable (Jacobi-like) extensions 
of the characters using affine Lie algebras associated with a given 
tenable solution; these impose further positivity constraints. The 
difficulty there lies in identifying the affine Lie algebra appropriate 
to a tenable solution. We are studying this aspect as well~\cite{tenable}.

\acknowledgments S.G. is supported by ANRF MATRICS grant number 
ANRF/ARGM/2025/00849/MTR. A.J. has been supported by an appointment to 
the JRG program at APCTP through the Science and Technology Promotion 
fund and Lottery fund of the Korean government. A.S. is supported by a 
grant from the Women Leading IITM program. We acknowledge the use of 
Claude Opus 5 (Anthropic) to get an efficient implementation of Virasoro 
inner products in Mathematica as well as computing the unitary F-matrix 
mentioned in Section \ref{sec:conj}. Both results were independently 
checked by us.

\appendix
\section{Notation}
\begin{center}
\begin{tabular}{c|l}
Symbol     &  Meaning \\ \midrule
  $\zeta_n$   &  Root of unity $\exp(2\pi i/n)$ \\
  $\mathcal{M}(p,p')$ & Virasoro minimal model with central charge $c=1-\frac{6}{pp'}$ \\
  $\langle \phi_1|\phi_2(1)\phi_3(z)|\phi_4(0)\rangle$ &
  SL$(2,\mathbb{C})$-fixed four-point function\\
  $z$ &  SL$(2,\mathbb{C})$-fixed conformal cross-ratio\\
  $\langle 1234\rangle_p(z) $ & Chiral conformal block for primary $\phi_p$ \\
  $[1234]_p(z)$ & Core conformal block for primary $\phi_p$\\
  ${\mathcal{M}^{(p)}_{\lambda\lambda'}}$ & Shapovalov matrix element between two partitions $\lambda$ and $\lambda'$ \\ & ~~for primary $\phi_p$   at level $n=|\lambda|=|\lambda'|$ \\
  $\rho_p^{(\alpha)}$ & Exponents about the singular points $\alpha=0,1,\infty$ \\
  $N_{ij}^k$ & Fusion coefficients \\
  $c_{ij}^k$  & Three-point structure constants \\
  $F_{st}$, $F_{s't'}$, $F_{u'u}$ & Braiding F-matrices connecting various channels \\
  $\bar{F}_{st}$, $\bar{F}_{s't'}$, $\bar{F}_{u'u}$ &  Unitary F-matrices  \\
  $R_{ss'}$, $R_{tu'}$, $R_{t'u}$  & Braiding R-matrices \\
  \bottomrule
\end{tabular}
\end{center}

\section{The Shapovalov matrix at small levels}
\label{sec:shapmatrix}

At level two with basis $(2,1^2)$, we have
\[
\mathcal{M}_2 =\begin{pmatrix}
 4h + \tfrac{c}2 &  6h \\ 6h & 4h(2h+1)  
\end{pmatrix}
\]
At level three, in the basis $(3,2~1,1^3)$, we have
\[
\mathcal{M}_3 = \begin{pmatrix}
  6h+2c &   10h & 24h \\
  10h & h(8h+c+8) & 12h(3h+1)\\
  24h & 12h(3h+1) & 24h(h+1)(2h+1)
\end{pmatrix}
\]

  At level four, we have a five-dimensional basis $(4,3~1,2^2,2~1^2,1^4)$. 
  \[
 \mathcal{M}_4=   {\footnotesize\left(
\begin{smallmatrix}
     5 c+8 h & 14 h & 3 (c+8 h) & 36 h & 120 h \\
 14 h & 4 h (c+3 h+3) & 30 h & 20 h (2 h+1) & 48 h (4 h+1) \\
 3 (c+8 h) & 30 h & \frac{c^2}{2}+c (8 h+4)+32 h (h+1) & 6 h (c+8 h+8) & 72 h
   (3 h+2) \\
 36 h & 20 h (2 h+1) & 6 h (c+8 h+8) & 2 h \left(2 c h+c+16 h^2+58 h+16\right)
   & 48 h \left(6 h^2+7 h+2\right) \\
 120 h & 48 h (4 h+1) & 72 h (3 h+2) & 48 h \left(6 h^2+7 h+2\right) & 96 h
   \left(4 h^3+12 h^2+11 h+3\right) 
\end{smallmatrix}\right)
}
\]

\section{The BPZ equation}\label{bpzsection}

In this section, we will focus on the chiral four-point functions 
containing a null primary at level four.

It is possible to express the chiral four-point functions in terms of 
the chiral conformal blocks defined in \eqref{Fblockdef}, which, in 
turn, can be written in terms of the core conformal blocks as in 
\eqref{coreblockdef}. For minimal models, the chiral four-point 
functions satisfy the BPZ equations. Using the BPZ equations, it is 
possible to derive an ordinary differential equation (ODE) obeyed by the 
core blocks. Upon the identification of the cross-ratio with (a sutiable 
function of) the modular parameter $\tau$, these ODEs become modular 
linear differential equations (MLDE). The order of each of these 
differential equations is determined by the lowest null level of the 
primary fields appearing in the correlation function under 
consideration. These equations are well known for orders two and three. 
Denoting $\mathcal{F}(z)$ to be the core block vector, the ODE at order 
two reads as~\cite{DiFrancesco:1997nk}:

\begin{eqnarray}\label{bpz2}
\left[\frac{d^2}{dz^2} + \frac{A^{(1)}(z)}{z(z-1)}  \frac{d}{dz} + \frac{A^{(2)}(z)}{z^2(z-1)^2} \right] \mathcal{F}(z) = 0
\end{eqnarray}
where the polynomials $A^{(1)}(z)$ and $A^{(2)}(z)$ read as 
\begin{align}
&A^{(1)}(z) = - (2\mu_{12}+t) + 2 (\upmu+t)z, A^{(2)}(z) = (\mu_{12})_2 + t(\mu_{12}-h_2) \cr &
- \big(2\mu_{12}(\upmu-1) + t(2\mu_{12}-\mu_{24}-2h_2)\big) z + \big((\upmu)_2 +t(\upmu-h/3)\big)z^2 ,
\end{align}
with $t= 2(2h_1+1)/3$. Here, for easy reading, we have introduced the notation $\upmu=\mu_{12}+\mu_{14}$ and used  the falling factorial symbol
$$ (x)_n = \prod_{k=0}^{(n-1)} (x-k) . $$

At order three, we have 
\begin{eqnarray}\label{bpz3}
\left[\frac{d^3}{dz^3} + \frac{B^{(1)}(z)}{z(z-1)} \frac{d^2}{dz^2} + \frac{B^{(2)}(z)}{z^2(z-1)^2} \frac{d}{dz}
+\frac{B^{(3)}(z)}{z^3(z-1)^3}  \right] \mathcal{F}(z) = 0.
\end{eqnarray}
The polynomials $B^{(k)}(z)$ are given by
$$ 
B^{(k)}(z) = \frac{1}{\tilde b_1} \sum_{p=1}^3\sum_{q=1}^{k+1} \tilde 
b_p B^{(k)}_{pq} z^{q-1} , 
$$
with the coefficients $B^{(k)}_{pq}$ having the values
\begin{align}
&B^{(1)}_{1q} = 0 , B^{(1)}_{21} = 1, B^{(1)}_{22} = -2, B^{(1)}_{31} = 3\mu_{12}, B^{(1)}_{32} = 2\upmu, 
B^{(2)}_{11} = 1, B^{(2)}_{12} = -3, \cr &
B^{(2)}_{13} = 3, B^{(2)}_{21} = h_2 - 2\mu_{12},
B^{(2)}_{22} = 5\mu_{12} - h_1 - h_2+1,
B^{(2)}_{23} = (h/3) - 3\upmu - 1, \cr &
B^{(2)}_{31} = 3 (\mu_{12})_2, 
B^{(2)}_{32} = - 6\mu_{12}(\upmu-1),
B^{(2)}_{33} = 3\big((\mu_{12})_2 + (\mu_{14})_2 + 2\mu_{12}\mu_{14}\big) \ .
\end{align}
for $k=1$ and $2$. For $k=3$ we have 
\begin{align}
&B^{(3)}_{11} = 2h_2 -\mu_{12},
B^{(3)}_{12} =  3\mu_{12}-\mu_{24}-6h_2, 
B^{(3)}_{13} = 3(h_1+2h_2-h_4), \cr &
B^{(3)}_{14} = -(2h_1+h_2+h_4),
B^{(3)}_{21} = (\mu_{12})_2 - h_2 \mu_{12},
B^{(3)}_{22} = \mu_{12}(h_1+2h_2) -3(\mu_{12})_2, \cr &
B^{(3)}_{23} = 3(\mu_{12})_2 - \mu_{12}(2h_1+h_2+h_4) - \mu_{14}(h_1+h_2),
B^{(3)}_{24} = - (\upmu)_2 + \upmu h/3, \cr &
B^{(3)}_{31} = - (\mu_{12})_3,
B^{(3)}_{32} =  3\big((\mu_{12})_3 + (\mu_{12})_2\mu_{14}\big),
B^{(3)}_{33} = - 3\big((\mu_{12})_3 + \mu_{12}(\mu_{14})_2 \cr & 
+ 2 \mu_{14}(\mu_{12})_2\big),
B^{(3)}_{34} =   (\mu_{12})_3 + (\mu_{14})_3 + 3 \mu_{12} (\mu_{14})_2 + 3 \mu_{14} (\mu_{12})_2\big) \ .
\end{align}
In addition, we have $\tilde b_1 = \tilde b_3 h_1(h_1+1), \tilde b_2 = - 
2 \tilde b_3 (h_1+1)$ for arbitrary $\tilde b_3$.

We now derive the ODE satisfied by the core blocks containing a null 
primary at level four. Let $|h_1\rangle$ be a primary state with weight 
$h_1$.  We choose the following basis of states at level four in the 
corresponding Verma module:
\begin{equation}
\left\{ L_{-4} |h_1\rangle, L_{-3}L_{-1}|h_1\rangle, L_{-2}^2|h_1\rangle,
L_{-2}L_{-1}^2|h_1\rangle, L_{-1}^4|h_1\rangle
\right\}
\end{equation}
A null state $|\chi\rangle$ at level four
\begin{equation}
|\chi\rangle = b_1 L_{-4} |h_1\rangle + b_2 L_{-3}L_{-1}|h_1\rangle
+ b_3 L_{-2}^2|h_1\rangle + b_4 L_{-2}L_{-1}^2|h_1\rangle + b_5 L_{-1}^4|h_1\rangle
\end{equation}
must satisfy $L_n|\chi\rangle = 0$ for all $n>0$. This condition is 
trivially satisfied by all $n\geq 5$. However, for $4\geq n\geq 1$, 
this imposes nontrivial relations among the coefficients $b_i$.
We find
\begin{align}
&5 b_1 + 2 h_1 b_2 + 3 b_3 = 0, \cr
&3 b_4 + 4 b_5 (2h_1 +3) = 0, \cr
&4 b_2 + 6 b_3 + 2 (2h_1+1) b_4 = 0,\cr
&6 b_1 + b_3 (8 h_1 + 8 +c) + 6 h_1 b_4 = 0, \cr
&5 b_2 + b_4 \left(4h_1 + 8 + \frac{c}{2}\right) + 12 b_5 (3 h_1+2) = 0, \cr
&7 b_1 + 2 b_2 (3h_1+3+c) + 15 b_3+ 10 b_4(2 h_1+1) + 24 b_5(4 h_1 + 1) = 0, \cr
&b_1 (8h_1+5c) + 14 h_1 b_2 + 6 b_3 \left(4h_1+\frac{c}{2}\right) + 36 h_1 b_4 + 120 h_1 b_5 = 0.
\end{align}
These equations admit solutions for the coefficients $b_i$ provided the 
weight $h_1$ of the primary state is related to the central charge $c$ 
by
$$ (8h_1+c-1)(16h_1 + 10 h_1c-82h_1+15c+66) = 0. $$
The possible values of $h_1$ are 
\begin{equation}
h_1 = \frac{1}{8}(1-c),
\end{equation}
and 
\begin{equation}
h_1 = \frac{1}{16} \left((41-5c)\pm\sqrt{(c-1)(c-25)}\right),
\end{equation}
which is consistent with the Kac formula.
For $h_1=(1-c)/8$, we have the values 
$$
b_2 = \frac{4h_1-3}{6h_1}b_1,\ 
b_3 = - \frac{4}{9}(h_1+3) b_1, \ 
b_4 = \frac{2h_1+3}{3h_1} b_1,\ 
b_5 = - \frac{b_1}{4h_1}.
$$
For $h_1 = \frac{1}{16} \left((41-5c)\pm\sqrt{(c-1)(c-25)}\right)$, 
we have 
\begin{eqnarray}
b_2 &=& - \frac{80h_1-5}{32 h_1^2+16h_1+27}b_1 ,\cr
b_3 &=& - \frac{15(2h_1+3)}{32 h_1^2+16h_1+27}b_1 ,\cr
b_4 &=& - \frac{125}{32 h_1^2+16h_1+27}b_1 ,\cr
b_5 &=& - \frac{375}{4(2h_1+3)(32 h_1^2+16h_1+27)}b_1.
\end{eqnarray}

The BPZ equation for the chiral blocks involving the field $\phi_1(z_1)$ 
is given by
\begin{equation}
\big(b_1{\mathcal L}_{-4}  + b_2 {\mathcal L}_{-3}{\mathcal L}_{-1}
+ b_3 {\mathcal L}_{-2}^2 + b_4 {\mathcal L}_{-2}{\mathcal L}_{-1}^2 + b_5 {\mathcal L}_{-1}^4\big){\mathcal F}_ {\phi_1,\phi_2,\phi_3,\phi_4}= 0 .   
\label{nbpz}
\end{equation}
where, the operators ${\mathcal L}_{-r}$ are defined as 
\begin{equation}
{\mathcal L}_{-r} = \sum_{i=2}^4 \left(\frac{(r-1)h_i}{z^r_{i1}} - \frac{1}{{\phantom{abc}}z^{(r-1)}_{i1}} \partial_{z_i}\right).
\label{difopl}
\end{equation}
We will consider the action of these operators on the chiral block 
${\mathcal F}$ and take the limits $z_1\rightarrow z, z_2\rightarrow 0, 
z_3\rightarrow\infty, z_4\rightarrow 1$ to obtain the desired ODE for 
the core block. Note that, in this limit,
$$
\partial_{z_1}^r {\mathcal F}_ 
{\phi_1,\phi_2,\phi_3,\phi_4}(z_1,z_2,z_3,z_4)= \left(\prod_{1\leq 
j,k\leq 4}z_{jk}^{\mu_{jk}} \right) \eth^r \mathcal{F}(z),
$$
where the action of the derivative operator $\eth$ is given by
\begin{eqnarray}
\eth \mathcal{F}(z) &= & \dfrac{d\mathcal{F}}{dz} + \frac{\upmu}{z(z-1)}  \left( z - \frac{\mu_{12}}{\upmu}\right) \mathcal{F}(z), \cr
\eth^2 \mathcal{F}(z) &=& \dfrac{d^2\mathcal{F}}{dz^2} +  \frac{2\upmu}{z(z-1)}  \left( z - \frac{\mu_{12}}{\upmu}\right) \dfrac{d\mathcal{F}}{dz} \cr
&+&  \frac{(\upmu)_2}{z^2(z-1)^2} \left( z^2 - \frac{2\mu_{12} }{\upmu} z + \frac{(\mu_{12})_2}{(\upmu)_2}\right)  \mathcal{F}(z) ,\cr
\eth^3 \mathcal{F}(z) &=& \dfrac{d^3\mathcal{F}}{dz^3} +  \frac{3\upmu}{z(z-1)} \left( z - \frac{\mu_{12}}{\upmu}\right)\dfrac{d^2\mathcal{F}}{dz^2}  \cr
&+&  \frac{3(\upmu)_2}{z^2(z-1)^2} \left( z^2 - \frac{2\mu_{12} }{\upmu} z + \frac{(\mu_{12})_2}{(\upmu)_2}\right) \dfrac{d\mathcal{F}}{dz} \cr
&+& \frac{(\upmu)_3}{z^3(z-1)^3} \left(z^3 - \frac{3\mu_{12}}{\upmu} z^2  + \frac{3 (\mu_{12})_2}{(\upmu)_2}  z
- \frac{(\mu_{12})_3}{(\upmu)_3 } \right) \mathcal{F}(z) ,\cr
\eth^4 \mathcal{F}(z) &=& \dfrac{d^4\mathcal{F}}{dz^4} +  \frac{4\upmu}{z(z-1)} \left( z - \frac{\mu_{12}}{\upmu}\right) \dfrac{d^3\mathcal{F}}{dz^3}  \cr
&+& \frac{6(\upmu)_2}{z^2(z-1)^2} \left( z^2 - \frac{2\mu_{12} }{\upmu} z + \frac{(\mu_{12})_2}{(\upmu)_2}\right) \dfrac{d^2\mathcal{F}}{dz^2}  \cr
&+& \frac{4(\upmu)_3 }{z^3(z-1)^3} \left(z^3 - \frac{3\mu_{12}}{\upmu} z^2  + \frac{3 (\mu_{12})_2}{(\upmu)_2}  z
- \frac{(\mu_{12})_3}{(\upmu)_3 } \right) \dfrac{d\mathcal{F}}{dz} \cr
&+& \frac{(\upmu)_4}{z^4(z-1)^4} \left( z^4 - \frac{4\mu_{12}}{\upmu} z^3
+ \frac{6(\mu_{12})_2}{(\upmu)_2} z^2 - \frac{4 (\mu_{12})_3}{(\upmu)_3} z + \frac{(\mu_{12})_4}{(\upmu)_4}
\right)\mathcal{F}(z). \nonumber
\end{eqnarray}

Using the above relations along with the differential operator 
\eqref{difopl} in the BPZ equation \eqref{nbpz}, we obtain the following 
ODE:
\begin{equation}\label{bpz4}
 \dfrac{d^4\mathcal{F}}{dz^4} + \sum_{k=1}^4  \frac{b_k  C^{(k)}(z)}{z^k (z-1)^k}\dfrac{d^{4-k}\mathcal{F}}{dz^{4-k}} =0,
\end{equation}
where the polynomials $C^{(k)}(z)$ are define by 
$$ C^{(k)}(z) = \frac{1}{b_1}\sum_{p=1}^5\sum_{q=1}^{k+1} C^{(k)}_{pq}z^{q-1}. $$
For $k=1,2$, the coefficients $C^{(k)}_{pq}$ are given as:
\begin{equation}
C^{(1)}_{1q} = C^{(1)}_{2q} = C^{(1)}_{3q} = 0, C^{(1)}_{41} = 1, C^{(1)}_{42} = -2, 
C^{(1)}_{51} = -4\mu_{12}, C^{(1)}_{52} = 4\upmu ,
\end{equation}
and
\begin{align}
&C^{(2)}_{1q} = 0, C^{(2)}_{21} = 1, C^{(2)}_{22} =-3, C^{(2)}_{23} = 3, C^{(2)}_{31} =1, C^{(2)}_{32} = -4, C^{(2)}_{33} = 4\cr &
C^{(2)}_{41} = h_2-3\mu_{12}, 
C^{(2)}_{42} = 8\mu_{12} + 2\mu_{14} - \mu_{24} - 2h_2 +2,
C^{(2)}_{43} = h/3 - 5\upmu - 2, \cr &
C^{(2)}_{51} = 6(\mu_{12})_2, 
C^{(2)}_{52} =-12(\mu_{12})_2 - 12\mu_{12}\mu_{14}, 
C^{(2)}_{53} = 6 (\upmu)_2 \ .
\end{align}
For $k=3$ we have 
\begin{align}
&C^{(3)}_{11} = 1, 
C^{(3)}_{12} = -4,
C^{(3)}_{13} = 6,
C^{(3)}_{14} = -4,
C^{(3)}_{21} = 2(h_2-\mu_{12}), \cr &
C^{(3)}_{22} = 7 \mu_{12} + \mu_{14} - \mu_{24} +1,
C^{(3)}_{23} = 3(h_1+h_2 - 3\mu_{12} -1), \cr &
C^{(3)}_{24} = 2(2\mu_{12}+\mu_{14}-h_1-h_4+1), 
C^{(3)}_{31} = 2h_2-2\mu_{12} - 1 , \cr &
C^{(3)}_{32} = 2(4\mu_{12} - \mu_{24} - 4 h_2 - 1),
C^{(3)}_{33} = 2(3h_1+5h_2-2h_4-2\mu_{12}-\mu_{14}), \cr &
C^{(3)}_{34} = 4(\upmu-h/3), 
C^{(3)}_{41} = 3(\mu_{12})_2 - 2h_2\mu_{12},
C^{(3)}_{42} = 2\big(\mu_{12}(h_1+2h_2-\mu_{14}) \cr &
+h_2\mu_{14} - 4(\mu_{12})_2 - \mu_{12}^2\big), 
C^{(3)}_{43} = 11(\mu_{12})_2 - (\mu_{14})_2 - 2(h_1+h_2)\mu_{14} \cr &
+ 2 \mu_{12} (4\mu_{14}-2h_1-h_2-h_4+2), 
C^{(3)}_{44} = -4(\upmu)_2 - 2\upmu +2h\upmu/3, \cr &
C^{(3)}_{51} = -4(\mu_{12})_3 , 
C^{(3)}_{52} = 12\big(\mu_{14}(\mu_{12})_2 + (\mu_{12})_3\big),
C^{(3)}_{53} = -12(\upmu)_3\mu_{12}/\upmu, \cr &
C^{(3)}_{54} = 4(\upmu)_3 \ .
\end{align}
Finally, for $k=4$ we have the expressions
\begin{align}
&C^{(4)}_{11} = 3h_2-\mu_{12}, C^{(4)}_{12} = 4\mu_{12} - \mu_{24} - 12 h_2, C^{(4)}_{13} = 2(2\mu_{24} - 3\mu_{12}+ 9h_2), \cr
&C^{(4)}_{14} = 2(2\mu_{12}-3\mu_{24}-6h_2), C^{(4)}_{15} = \mu_{24}+2h-2h_3, C^{(4)}_{21} = \mu_{12}(\mu_{12}-2h_2-1), \cr
&C^{(4)}_{22} = \mu_{12}(h_1+7h_2-4\mu_{12}+4) + 2h_2\mu_{14} - \mu_{14} (\mu_{24}+6h_2) - \mu_{12}(4h_1+9h_2\cr
&-h_4-5\mu_{12}+6), C^{(4)}_{23} = 3\mu_{14} (\mu_{24}+2h_2) + \mu_{12}(5h_1+4h_2+h_4-3\mu_{12}+4), \cr
&C^{(4)}_{25} = (\upmu)_2 - 2h\upmu/3, C^{(4)}_{31} = (\mu_{12}-h_2)(\mu_{12}-h_2-2), 
C^{(4)}_{32} = 2\mu_{12}(\mu_{24}-2\mu_{12} \cr
&+4h_2+4) - 2h_2(\mu_{24} + 2h_2+4)-\mu_{24}, C^{(4)}_{33} = \mu_{24}^2 + 3 \mu_{24}(2h_2-2\mu_{13}+1) \cr
&+ 2(h_2-\mu_{12})\big(3(h_2-\mu_{12})+h_4-\mu_{14}+6\big),  C^{(4)}_{34} = 2(\mu_{12}-h_1-h_4-2)(\mu_{24}\cr
&-2\mu_{12}+2h_2), C^{(4)}_{35} =(\mu_{12}-h_1-h_4)(\mu_{12}-h_1-h_4-2), C^{(4)}_{41} = (\mu_{12})_3 - h_2(\mu_{12})_2, \cr
&C^{(4)}_{42} = (\mu_{12})_2(h_1-5h_2-\mu_{14}) - 4 (\mu_{12})_3 - 2 h_2\mu_{12}\mu_{14}, C^{(4)}_{43} = -6(\mu_{12})_3 \cr 
&+ (3h_1+3h_2+h_4-4\mu_{14})(\mu_{12})_2 +h_2 (\mu_{14})_2+\mu_{12}\mu_{14}(\mu_{24}+h_1+5h_2+1), \cr
&C^{(4)}_{44} = (\upmu-1)\big(4(\mu_{12})_2 + \mu_{12}(\mu_{24} - 4 h_1 - 2 h_4 - 4) - \mu_{14}(\mu_{24}+ 2 h_2)\big), \cr &
C^{(4)}_{45} =(h/3)(\upmu)_2 - (\upmu)_3, C^{(4)}_{51} = (\mu_{12})_4,  C^{(4)}_{52} = -4 (\mu_{12})_3 (\upmu)_4/(\upmu)_3, \cr
&C^{(4)}_{53} = 6(\mu_{12})_2(\upmu)_4/(\upmu)_2, C^{(4)}_{54} = - 4\mu_{12} (\upmu)_4/\upmu,
C^{(4)}_{55} = (\upmu)_4 \ .
\end{align}


\bibliographystyle{JHEP}
\bibliography{master}

\end{document}